\documentclass[12pt]{article}

\usepackage{amsfonts}
\usepackage{amsmath}
\usepackage{amsthm}
\usepackage{cite}
\usepackage{epsfig}
\usepackage{latexsym}
\usepackage{paralist}
\usepackage{fancyhdr}
\usepackage{graphicx}
\numberwithin{equation}{section}
\usepackage[vcentermath]{youngtab}
\usepackage{young}
\usepackage{ytableau}
\usepackage{etex}
\usepackage{braket}
\usepackage{float}

\usepackage{pict2e}   
\usepackage{animate}  

\usepackage{multirow}
\usepackage{bigdelim}
\usepackage{fancybox}
\usepackage{cals}

\def\be{\begin{equation}}
\def\ee{\end{equation}}
\def\bea{\begin{eqnarray}}
\def\eea{\end{eqnarray}}

\renewcommand{\thefootnote}{\fnsymbol{footnote}}

\begin{document}

\hfuzz=100pt
\title{{\Large \bf{Duality and Confinement \\in 3d $\mathcal{N}=2$ ``chiral'' $SU(N)$ gauge theories} }}
\date{}
\author{ Keita Nii$^a$\footnote{nii@itp.unibe.ch}
}
\date{\today}

\maketitle

\thispagestyle{fancy}
\cfoot{}
\renewcommand{\headrulewidth}{0.0pt}

\vspace*{-1cm}
\begin{center}
$^{a}${{\it Albert Einstein Center for Fundamental Physics }}
\\{{\it Institute for Theoretical Physics
}}
\\ {{\it University of Bern}}  
\\{{\it  Sidlerstrasse 5, CH-3012 Bern, Switzerland}}

\end{center}

\begin{abstract}
We study low-energy dynamics of three-dimensional $\mathcal{N}=2$ $SU(N)$ ``chiral'' gauge theories with $F$ fundamental and $\bar{F}$ anti-fundamental matters without a Chern-Simons term. Compared to a naive semi-classical analysis of the Coulomb branch, its quantum structure is highly richer than expected due to so-called ``dressed'' Coulomb branch (monopole) operators. We propose dualities and confinement phases for the ``chiral'' $SU(N)$ theories. The theories with $N>F > \bar{F}$ exhibit spontaneous supersymmetry breaking. The very many Coulomb branch operators generally remain exactly massless and are non-trivially mapped under the dualities. Some dualities lead to a novel duality between $SU(N)$ and $USp(2 \tilde{N})$ theories. For the 3d $\mathcal{N}=2$ $SU(2)$ gauge theory with $2F$ doublets, there are generally $F+2$ ``chiral'' and ``non-chiral'' dual descriptions. 
\end{abstract}

\renewcommand{\thefootnote}{\arabic{footnote}}
\setcounter{footnote}{0}

\newpage
\tableofcontents 
\clearpage

\newpage

\section{Introduction}
Asymptotically-free gauge theories exhibit various phases depending on gauge groups, matter contents, space-time dimensions and so on. It is generally difficult to analytically study the low-energy dynamics because those theories are often strongly-coupled and non-perturbative. 
Duality is a very powerful tool for investigating the dynamics which is not accessible by a perturbation theory \cite{Seiberg:1994pq}. Duality gives us other descriptions of the theories, which are more tractable than the original theories. Supersymmetry is also very powerful and allows us to exactly obtain the non-perturbative information \cite{Seiberg:1994bz}.

Recently, a deeper understanding of the three-dimensional $\mathcal{N}=2$ supersymmetric gauge theories was gained a lot. In particular, the 3d Seiberg dualities were derived from the corresponding 4d dualities \cite{Aharony:2013dha, Aharony:2013kma}. In this derivation, it was important to correctly understand the Coulomb branch (monopole) operators and the non-perturbative superpotential coming from a circle compactification of the 4d theories. The analysis of the Coulomb branch was developed also for theories with various matters and various gauge groups \cite{Csaki:2014cwa,Amariti:2015kha,Nii:2016jzi,Nii:2018erm,Nii:2017npz,Nii:2018tnd}. 
For 3d ``chiral'' theories with unequal numbers of (anti-)fundamental matters, the dualities were discussed in the $U(N)$ cases \cite{Benini:2011mf}. For the $SU(N)$ cases, a completely chiral theory with no anti-fundamental matter was discussed in \cite{Aharony:2013dha}. In \cite{Intriligator:2013lca}, various chiral theories were extensively studied for the $U(1)$ and $SU(2)$\footnote{\cite{Intriligator:2013lca} considered the 3d $\mathcal{N}=2$ $SU(2)$ Chern-Simons-Matter theory with odd number of fundamentals.} cases with and without a Chern-Simons coupling.

However, the quantum structure of the Coulomb branch for the $SU(N)$ ``chiral'' gauge theory with $F$ fundamental and $\bar{F} (< F)$ anti-fundamental matters\footnote{We simply denote this theory as $SU(N)$ with $(F,\bar{F})$ (anti-)fundamentals.} has been less understood until now. Although the ``chiral'' $SU(N)$ theories were also studied in \cite{Intriligator:2013lca}, its analysis was semi-classical and discussed only a very limited class of the Coulomb branch which corresponds to the breaking $SU(N) \rightarrow SU(N-2) \times U(1) \times U(1)$. Since there are so many other classical Coulomb branches, this analysis \cite{Intriligator:2013lca} would be incomplete and miss some important quantum aspects of the ``chiral'' theories.
For instance, \cite{Intriligator:2013lca} claims that there is no s-confining phase for the 3d ``chiral'' $SU(N)$ theories since we cannot contract the flavor indices of the meson operator and then a confining superpotential is not available.  \cite{Intriligator:2013lca} also claims that the supersymmetry will be broken for $N > \bar{F}$ without mentioning the allowed region of $F$. They concluded these two statements only from the semi-classical treatment of the Higgs branch and one particular Coulomb branch. By carefully studying various Coulomb branches, we will find that the first statement is incorrect and the second statement should be refined into $N>F>\bar{F}$.

In this paper, we will consider the three-dimensional $\mathcal{N}=2$ $SU(N)$ gauge theory with $(F,\bar{F})$ (anti-)fundamentals with $F > \bar{F}$. We will claim that the very many classical Coulomb branches can survive quantum corrections and remain exactly massless in the ``chiral'' $SU(N)$ theories. This situation is very different from the vector-like theory with $F=\bar{F}$, where only a single Coulomb branch remains flat. Naively, one might consider that many Coulomb branches are not gauge invariant and they cannot be moduli coordinates due to the ``chirality''. However, we can construct the so-called ``dressed'' monopole operators by appropriately combining the bare Coulomb branch and matter operators, which are gauge-invariant and quantum-mechanically flat. 
By taking into account these ``dressed'' operators, we will present various phases: For $F>N+1$, we will propose a dual description in terms of the $SU(F-N)$ gauge theory. For $F=N+1$, we will find a confinement phase where the low-energy degrees of freedom are governed by a cubic superpotential. For $N>F>\bar{F}$, the theory exhibits a spontaneous supersymmetry breaking. 
In contrast to the ``chiral'' $U(N)$ duality, the ``chiral'' $SU(N)$ duality gives rise to an additional problem of matching the baryon operators under the duality. 
We will demonstrate that the anti-baryonic operator is mapped to one of the dressed Coulomb branch operators. 
By focusing on the duality with the particular matter contents, we will also propose a novel duality between the chiral $SU(N)$ and $USp(2 \tilde{N})$ theories.

The rest of the paper is organized as follows. In Section 2, we will introduce a set of the Coulomb branch coordinates which appear in the 3d $\mathcal{N}=2$ ``chiral'' $SU(N)$ theory. We divide the argument into $SU(2N)$ and $SU(2N+1)$ cases.
In Section 3, we propose the dualities of the ``chiral'' $SU(N)$ theory. Since the precise structure of the Coulomb branch depends on $F, \bar{F}$ and $N$, we will show dualities case-by-case in the subsequent subsections. The first two cases lead to the novel dualities between $SU(N)$ and $USp(2 \tilde{N})$ theories.
In Section 4, we study the confinement phases in the 3d ``chiral'' $SU(N)$ theories. By deforming the confinement phases, we can find supersymmetry-breaking phases and quantum-deformed moduli spaces for particular matter contents.
In Section 5, we will summarize our findings and discuss the future directions.
In Appendix, we will show an example of the various chiral and non-chiral dualities for the 3d $SU(2)$ gauge theory with $2F$ fundamental matters, especially focusing on the $F=3$ case.

\section{Coulomb branch in ``chiral'' $SU(N)$ theories}

In this section, we will introduce a ``typical'' set of the Coulomb branch coordinates for the 3d $\mathcal{N}=2$ $SU(N)$ gauge theories with chiral matter contents. By carefully studying the gauge invariance of these coordinates and by taking into account induced Chern-Simons couplings for unbroken gauge groups, we can find which coordinates remain exactly massless at a quantum level. Note that in this section we will only consider the ``typical'' Coulomb branch in the 3d chiral $SU(N)$ theories. Then, there might be additional Coulomb brach operators which could appear in case of particular matter contents and the ranks of the gauge group. In some cases, the Coulomb branch operators which will be introduced below are not minimal-monopole-creating operators and we must take $n$-th root of these operators in those cases. However, the understanding of these ``typical'' Coulomb branches will be sufficient and strong enough to understand various confinement phases and dualities.

\subsection{$SU(2N)$ with $(F,\bar{F})$}

We start with the Coulomb branch of the 3d $\mathcal{N}=2$ $SU(2N)$ gauge theories with $(F,\bar{F})$ (anti-)fundamentals. The matter contents and their representations are summarized in Table \ref{MCSU2N} below. Notice that, due to a parity anomaly, $F \pm \bar{F}$ should be even since we do not introduce a tree-level Chern-Simons term to the gauge interaction. 

\begin{table}[H]\caption{Matter contents of $SU(2N)$ with $(F,\bar{F})$} 
\begin{center}
\scalebox{1}{
  \begin{tabular}{|c||c||c|c|c|c|c| } \hline
  &$SU(2N)$&$SU(F)$&$SU(\bar{F})$&$U(1)$&$U(1)$&$U(1)_R$  \\ \hline
 $Q$ & ${\tiny \yng(1)}$ & ${\tiny \yng(1)}$&1&1&0&$r$ \\
$\tilde{Q}$  &${\tiny \overline{\yng(1)}}$&1& ${\tiny \yng(1)}$&$0$&1& $\bar{r}$ \\  \hline
  \end{tabular}}
  \end{center}\label{MCSU2N}
\end{table}
The Coulomb branch (or Coulomb moduli space) is a flat direction of the adjoint scalar fields which come from the $SU(2N)$ vector multiplet.  Typically, we will consider the following Coulomb branch which breaks the gauge group as
\begin{align}
SU(2N)  &\rightarrow SU(N-a) \times SU(N-a) \times SU(2a) \times U(1)_1  \times U(1)_2 \\
U(1)_1 &\sim \mathrm{diag}(1,\cdots,1, 0,\cdots,0,-1,\cdots,-1) \\
U(1)_2 &\sim \mathrm{diag} \left( 1,\cdots,1 , -\frac{N-a}{a} ,\cdots,-\frac{N-a}{a},1,\cdots,1  \right) \\
{\tiny \yng(1)} & \rightarrow ({\tiny \yng(1)},\cdot,\cdot)_{1,1}+(\cdot,{\tiny \yng(1)},\cdot)_{-1,1}+(\cdot,\cdot,{\tiny \yng(1)})_{0,-\frac{N-a}{a}} \\
\overline{{\tiny \yng(1)}} & \rightarrow (\overline{{\tiny \yng(1)}},\cdot,\cdot)_{-1,-1}+(\cdot, \overline{{\tiny \yng(1)}},\cdot)_{1,-1}+(\cdot,\cdot, \overline{{\tiny \yng(1)}})_{0,\frac{N-a}{a}},
\end{align}
where $a=1,\cdots,N-1$. For $a=0$, the breaking pattern is slightly modified to
\begin{align}
SU(2N)  &\rightarrow SU(N) \times SU(N)  \times U(1)_1 \\
{\tiny \yng(1)} & \rightarrow ({\tiny \yng(1)},\cdot)_{1} +(\cdot,{\tiny \yng(1)})_{-1}\\
\overline{{\tiny \yng(1)}} & \rightarrow ( \overline{{\tiny \yng(1)}},\cdot)_{-1} +(\cdot, \overline{{\tiny \yng(1)}})_{1}.
\end{align}
There are $N$ types of the Coulomb branches which are related with $U(1)_1$. The corresponding operators are denoted as $Y_a^{bare}$. For the theory with $(F,\bar{F})$ (anti-)fundamentals, along these directions, the effective Chern-Simons terms are introduced. The (anti-)fundamental fields are massive and integrated out, which results in an $SU(N-a)_{\frac{F-\bar{F}}{2}} \times SU(N-a)_{-\frac{F-\bar{F}}{2}} $ pure Chern-Simons theory.
If we have an $SU(M)_k$ CS theory with $0\le |k|<M$, there is no SUSY vacuum \cite{Intriligator:2013lca}. Therefore, we require $N-a \le \frac{F-\bar{F}}{2}$, which means $a$ must satisfy
\begin{align}
\frac{2N-(F-\bar{F})}{2}\le a.
\end{align}
Notice that the effective Chern-Simons term is not generated for $U(1)_1$ since the low-energy $U(1)_1$ theory is vector-like. That is why these Coulomb branches can be flat directions.

These Coulomb branch operators are generally not gauge invariant and charged under the $U(1)_2$ since the theory is ``chiral''. Therefore, we have to multiply it by the chiral superfields $(\cdot,\cdot, \overline{{\tiny \yng(1)}})_{0,\frac{N-a}{a}} \in \tilde{Q}$ which belong to an $SU(2a)$ anti-fundamental representation. Then, we have to construct the baryonic operators from them in order to make an $SU(2a)$ singlet, which requires 
\begin{align}
2a \le \bar{F} ~~~\mbox{for}~a \ne 0.
\end{align}
By collecting all the conditions for $a$, we find 
\begin{align}
 \frac{2N-(F-\bar{F})}{2} \le a  \le \frac{\bar{F}}{2},~~~0 \le a \le N-1
\end{align}
and the Coulomb branch operators exist for this range of $a$.
By computing the effective Chern-Simons term between $U(1)_1$ and $U(1)_2$, one can compute the $U(1)_2$ charge of the bare monopole operators $Y_a^{bare}$. By properly dressing the bare operators with $\tilde{Q}^{2a}$, the dressed Coulomb branches become
\begin{align}
Y_a^{dressed} \sim Y_a^{bare} (\tilde{Q}^{2a})^{\frac{F-\bar{F}}{2}}.
\end{align}
The charges of these operators can be computed from the mixed Chern-Simons terms as in Table \ref{CBSU2n} below.

\begin{table}[H]\caption{Charges of the monopole operators} 
\begin{center}
\scalebox{1}{
  \begin{tabular}{|c||c||c|c|c| } \hline
  &$U(1)_2$&$U(1)$&$U(1)$&$U(1)_R$  \\  \hline
  $Y_{0}^{bare}$&0&$-FN$&$-\bar{F}N$& \scriptsize  $-FN(r-1)-\bar{F}N(\bar{r}-1) -2N^2$ \\[3pt]
$Y_{a \neq 0}^{bare}$ &$-(F-\bar{F})(N-a)$ &$-F(N-a)$&$-\bar{F}(N-a)$&\scriptsize  \begin{tabular}{c} $-F(N-a)(r-1)-\bar{F}(N-a)(\bar{r}-1)$ \\ $-2(N-a)(N+a)$ \end{tabular}\\[7pt]
$Y_{a}^{dressed}$&0&$-F(N-a)$&$-\bar{F} N +aF$  &\scriptsize \begin{tabular}{c} $-F(N-a)r-(\bar{F} N -aF)\bar{r} $\\ $+(N-a)(F+\bar{F} -2(N+a))$ \end{tabular}  \\[7pt] \hline
  \end{tabular}}
  \end{center}\label{CBSU2n}
\end{table}

Finally, we must be careful of minimal monopole operators. Here we constructed the naive monopole operators and there might be more smaller operators whose magnetic charges are smaller than here. Since the $SU(N-a) \times SU(N-a)$ gauge symmetry is unbroken in addition to $U(1)_1$, the minimal $U(1)_1$ magnetic charge could be more smaller. In some cases, the corresponding (minimal) magnetic charge becomes $\frac{1}{N-a}$ and the minimal bare operators are the $(N-a)$-th root of the naive ones. However, for those minimal operators, we cannot make them gauge-invariant by using $\tilde{Q}^{2a}$ in many cases. If this is possible, we have to construct more minimal operators from the naive ones by taking a root. Thus, our analysis below will become case-by-case depending on the matter contents and the rank of the gauge group.

\subsection{$SU(2N+1)$ with $(F,\bar{F})$}
For completeness of our paper, we also study the ``typical'' Coulomb branch of the 3d $\mathcal{N}=2$ $SU(2N+1)$ theory with $(F,\bar{F})$ (anti-)fundamental matters, which needs a small modification of the previous analysis. The matter contents and the global symmetries are the same as the $SU(2N)$ case (see Table \ref{MCSUodd} below). 

\begin{table}[H]\caption{3d $\mathcal{N}=2$ $SU(2N+1)$ with $(F,\bar{F})$ (anti-)fundamentals} 
\begin{center}
\scalebox{1}{
  \begin{tabular}{|c||c||c|c|c|c|c| } \hline
  &$SU(2N+1)$&$SU(F)$&$SU(\bar{F})$&$U(1)$&$U(1)$&$U(1)_R$  \\ \hline
 $Q$ & ${\tiny \yng(1)}$ & ${\tiny \yng(1)}$&1&1&0&$r$ \\
$\tilde{Q}$  &${\tiny \overline{\yng(1)}}$&1& ${\tiny \yng(1)}$&$0$&1& $\bar{r}$ \\  \hline
  \end{tabular}}
  \end{center}\label{MCSUodd}
\end{table}

We consider the Coulomb moduli space whose expectation value breaks the gauge group as
\begin{align}
SU(2N+1)  &\rightarrow SU(N-a) \times SU(N-a) \times SU(2a+1) \times U(1)_1  \times U(1)_2,
\end{align}
which is labeled by $Y_a^{bare}$ $ (a=0,\cdots,N-1)$. The $SU(2a+1)$ factor is absent for $a=0$. Notice that there are two $U(1)$ factors even for $a=0$. This is different from the $SU(2N)$ case. $Y_0^{bare}$ is not gauge invariant now and would need ``dressing''. 
The (anti-)fundamental representations are decomposed into
\begin{align}
{\tiny \yng(1)} & \rightarrow ({\tiny \yng(1)},\cdot,\cdot)_{1,1}+(\cdot,{\tiny \yng(1)},\cdot)_{-1,1}+(\cdot,\cdot,{\tiny \yng(1)})_{0,-\frac{2(N-a)}{2a+1}} \\
\overline{{\tiny \yng(1)}} & \rightarrow (\overline{{\tiny \yng(1)}},\cdot,\cdot)_{-1,-1}+(\cdot, \overline{{\tiny \yng(1)}},\cdot)_{1,-1}+(\cdot,\cdot, \overline{{\tiny \yng(1)}})_{0,\frac{2(N-a)}{2a+1}},
\end{align}
where $(2a+1)$-dimensional representations become singlets for $a=0$ and are charged only under the $U(1)_2$. The effective Chern-Simons term between $U(1)_1$ and $U(1)_2$ is generated as
\begin{align}
k_{eff}^{U(1)_1 U(1)_2} =(F-\bar{F}) (N-a)
\end{align}
and the Coulomb branch operator $Y_a^{bare}$ is not gauge invariant (even for $a=0$). We can construct a gauge invariant composite by using $ (\cdot,\cdot, \overline{{\tiny \yng(1)}})_{0,\frac{2(N-a)}{2a+1}} \in \tilde{Q} $. The candidate of the dressed monopole operator is
\begin{align}
Y_a^{dressed} \sim Y_{a}^{bare} (\tilde{Q}^{2a+1})^{\frac{F-\bar{F}}{2}}.
\end{align}
The charges of these monopole operators are summarized in Table \ref{CBSUodd} below.
\begin{table}[H]\caption{Charges of monopole operators} 
\begin{center}
\scalebox{1}{
  \begin{tabular}{|c||c||c|c|c| } \hline
  &$U(1)_2$&$U(1)$&$U(1)$&$U(1)_R$  \\  \hline
$Y_{a}^{bare}$ & \footnotesize $-(F-\bar{F})(N-a)$ &$-F(N-a)$&$-\bar{F}(N-a)$&\tiny \begin{tabular}{c} $-F(N-a)(r-1)-\bar{F}(N-a)(\bar{r}-1)$ \\ $-2(N-a)(N+a+1)$ \end{tabular}\\
$Y_{a}^{dressed}$&1&$-F(N-a)$& \footnotesize  $-\bar{F} N +\frac{F-\bar{F}}{2}+aF$  &\tiny \begin{tabular}{c} $-F(N-a)r-(\bar{F} N -\frac{F-\bar{F}}{2}-aF)\bar{r} $\\ $+(N-a)(F+\bar{F} -2(N+a+1))$ \end{tabular}  \\ \hline
  \end{tabular}}
  \end{center}\label{CBSUodd}
\end{table}

Along the Coulomb branch, the matter fields are massive and integrated out. This introduces the Chern-Simons terms for the two $SU(N-a)$ groups. The resulting low-energy theory is an $SU(N-a)_{\frac{F-\bar{F}}{2}} \times SU(N-a)_{-\frac{F-\bar{F}}{2}} $ Chern-Simons theory. Since this theory must have a supersymmetric vacuum, it requires \cite{Intriligator:2013lca}
\begin{align}
N-a \le \frac{F -\bar{F}}{2}.
\end{align}
Furthermore, since we need to construct a baryonic operator $\tilde{Q}^{2a+1}$ in order to dress the bare monopoles, there is a further constraint, $2a+1 \le \bar{F}$. By collecting all the constraints for $a$, we find
\begin{align}
0\le a \le N-1,~~~\frac{2N-(F-\bar{F})}{2} \le a \le \frac{\bar{F}-1}{2}.
\end{align}
The Coulomb branch $Y_a^{dressed}$ exists only for this range of $a$.

In this subsection, we focused on the general structure of the Coulomb branch. When we will study a specific example in the next section, there are more-smaller monopole operators available in some cases. For instance, we can consider a monopole operator
\begin{align}
Y_{a}^{\substack{minimal \\ bare}} \sim (Y_{a}^{bare})^{\frac{1}{N-a}},
\end{align}
which has a smaller magnetic charge and is still not gauge invariant. In many cases, we cannot construct a gauge invariant composite from $Y_{a}^{\substack{minimal \\ bare}} $ and $\tilde{Q}^{2a+1}$. However, in some cases (special choices of $N,F,\bar{F}$ and $a$), we can do it. Because the correct choice of the bare monopoles depend on the matter contents, we have to consider the dualities case-by-case.

\section{Dualities}

Here, we propose a new duality for the 3d $\mathcal{N}=2$ $SU(N)$ gauge theory with $F$ fundamental matters $Q$ and $\bar{F}$ anti-fundamental matters $\tilde{Q}$. The following discussion assumes $F > \bar{F}$ without loss of generality. The proposed dual description is a 3d  $\mathcal{N}=2$ $SU(F-N)$ gauge theories with $F$ fundamental (dual) matters $q$, $\bar{F}$ anti-fundamental (dual) matters $\tilde{q}$ and a meson singlet $M$ with a following superpotential 
\begin{align}
W=M \tilde{q}q. \label{super}
\end{align}

In order to illustrate how the duality works, we restrict ourself to the case where the anti-baryonic operators cannot be constructed on the electric side. Furthermore, in order to explain a generic structure of the duality, let us focus on the 3d $\mathcal{N}=2$ $SU(2N)$ with $(2F,2\bar{F})$ (anti-)fundamentals. One can easily generalize our analysis to more generic cases. The proposed dual description is 
a 3d $\mathcal{N}=2$ $SU(2F-2N)$ with $(2F,2\bar{F})$ (anti-)fundamentals and a singlet $M$ with a superpotential \eqref{super}. 
The matter contents and their quantum numbers are summarized in Table \ref{generic}. 

\begin{table}[H]\caption{$SU(2N)$ with $(2F,2\bar{F})$ and its dual} 
\begin{center}
\scalebox{0.96}{
  \begin{tabular}{|c||c||c|c|c|c|c| } \hline
  &$SU(2N)$&$SU(2F)$&$SU(2\bar{F})$&$U(1)$&$U(1)$&$U(1)_R$  \\ \hline
 $Q$ & ${\tiny \yng(1)}$ & ${\tiny \yng(1)}$&1&1&0&$r$ \\
$\tilde{Q}$  &${\tiny \overline{\yng(1)}}$&1& ${\tiny \yng(1)}$&$0$&1& $\bar{r}$ \\  \hline
$M:=\tilde{Q}Q$&1&${\tiny \yng(1)}$& ${\tiny \yng(1)}$&1&1&$r+\bar{r}$\\ 
$B:=Q^{2N}$&1&\scriptsize \begin{tabular}{c}$2N$-th \\ antisymmetric\end{tabular}&1&$2N$&$0$&$2Nr$ \\ \hline \hline 
&$SU(2F-2N)$&$SU(2F)$&$SU(2 \bar{F})$&$U(1)$&$U(1)$&$U(1)_R$  \\ \hline
$q$& ${\tiny \yng(1)}$& ${\tiny \overline{\yng(1)}}$&1&$\frac{N}{F-N}$&$0$& $\frac{N}{F-N}r$ \\
$\tilde{q}$&${\tiny \overline{\yng(1)}}$&1 & $\overline{{\tiny \yng(1)}}$&$-\frac{F}{F-N}$&$-1$&$2-\frac{F}{F-N}r-\bar{r}$ \\
$M$&1&${\tiny \yng(1)}$ &${\tiny \yng(1)}$ &1&1&$r+\bar{r}$ \\\hline
$N:=\tilde{q}q$&1&${\tiny \overline{\yng(1)}}$ & $\overline{{\tiny \yng(1)}}$ &$-1$&$-1$&$2-r-\bar{r}$ \\ 
$b:=q^{2F-2N} \sim B$&1&\scriptsize \begin{tabular}{c}$2N$-th \\ antisymmetric\end{tabular}&1&$2N$&$0$&$2Nr$ \\ \hline
  \end{tabular}}
  \end{center}\label{generic}
\end{table}

The charge assignment of the fields in the dual description can be fixed from the superpotential and the matching of the baryon operator. 
The mesonic branch labeled by $M: = \tilde{Q} Q$ is mapped to a singlet $M$. The baryonic operator $B:=Q^{2N}$ is mapped to the baryon $b:=q^{2F-2N}$ constructed from the dual quarks $q$. When $\bar{F} \ge F-N $, we can construct the anti-baryonic operator on the magnetic side. One might consider that there is no counterpart of the anti-baryon on the electric side. However, as we will see in the next subsection, the anti-baryon corresponds to one of the Coulomb branch operators on the electric side.

The electric and magnetic theories are both UV-complete because the duality does not include a monopole superpotential. In addition, the proposed duality is very similar to the conventional 4d Seiberg duality \cite{Seiberg:1994pq}. Therefore, this 3d duality can straightforwardly pass several simple tests. By applying the duality twice, we can go back to the electric theory. The complex mass deformation on the electric side is mapped to the Higgsing of the dual gauge group and the duality is correctly preserved with reduction of $F$ and $\bar{F}$. The parity anomaly matching is also satisfied. The generic real mass deformation is now complicated and let us focus on one particular simple case. By weakly gauging the global $U(1)$ symmetries, all the electric matters can obtain the positive real masses and the electric theory flows to the $SU(2N)_{F+\bar{F}}$ SUSY CS theory. This flow is only possible for $F+\bar{F}  \ge 2N$ otherwise the supersymmetry is broken and we have to find a more complicated flow. On the magnetic side, the fundamental matters obtain positive real masses while the anti-fundamental matters obtain negative masses. The theory flows to $SU(2F-2N)_{F-\bar{F}}$ SUSY CS theory. This flow is possible for $F-\bar{F} \ge 2F-2N$. Therefore, only for $\bar{F}=2N-F$, these flows are simultaneously realized. In this case, the electric side becomes the $SU(2N)_{2N} $ CS theory while the magnetic side becomes the $SU(2F-2N)_{2F-2N}$ CS theory. The both sides have the same Witten index \cite{Witten:1999ds,Intriligator:2013lca} which is one. The most non-trivial part of this duality is the matching of the moduli operators and hence we focus on it in what follows.

Since the correct Coulomb branch operators will change case-by-case depending on $N,F,\bar{F}$ and $a$, we here only give a general rule of matching the Coulomb branch operators on both sides. The more precise matching (including the anti-baryon) will be shown in the following subsections. As we explained in the previous section, the Coulomb branch of the ``chiral'' theory is labeled by $Y_a$. The charges of these operator on both sides are summarized in Table \ref{CBmapping} below. Notice that the dressed operators are generally charged under the non-abelian flavor symmetry, but we omitted it for simplicity. 

\begin{table}[H]\caption{Coulomb branch operators} 
\begin{center}
\scalebox{0.63}{
  \begin{tabular}{|c||c||c|c|c| } \hline
  &$U(1)_2$&$U(1)$&$U(1)$&$U(1)_R$  \\ \hline
$Y_a^{bare}$&\footnotesize  $-(2F-2\bar{F})(N-a) $& \footnotesize $-2F(N-a)$&\footnotesize $-2\bar{F}(N-a)$&\tiny \begin{tabular}{c}
$-2F(N-a)r -2\bar{F}(N-a) \bar{r} $\\$+2F(N-a)+2\bar{F}(N-a)-2(N^2-a^2)$
\end{tabular}\\
\footnotesize $Y_a^{dressed}:=Y_a^{bare} \tilde{Q}^{2a \frac{2F-2\bar{F}}{2}}$&0& \footnotesize $-2F(N-a)$& \footnotesize $-2\bar{F}N+2aF$&\tiny \begin{tabular}{c}
$-2F(N-a)r -2\bar{F}(N-a) \bar{r} $\\$+2F(N-a)+2\bar{F}(N-a)-2(N^2-a^2)+a(2F-2\bar{F})\bar{r}$
\end{tabular} \\ \hline \hline 
&$U(1)_2$&$U(1)$&$U(1)$&$U(1)_R$  \\ \hline
$\tilde{Y}_a^{bare}$& \footnotesize $-(2F-2\bar{F})(F-N-a)$& \footnotesize $\frac{2F(F-N-a)(\bar{F}-N)}{F-N}$&\footnotesize $2 \bar{F}(F-N-a)$&\tiny \begin{tabular}{c}
$\frac{2F(F-N-a)(2\bar{F}-N)}{F-N} r +2 \bar{F}(F-N-a) \bar{r} $\\$+2F(F-N-a)-2\bar{F}(F-N-a)-2((F-N)^2-a^2)$
\end{tabular} \\ 
\footnotesize $\tilde{Y}_a^{dressed}:=\tilde{Y}_a^{bare} \tilde{q}^{2a\frac{2F-2\bar{F}}{2}}$&0& \footnotesize $-2F(N-\bar{F}+a)$& \footnotesize $-2\bar{F}N+2F(\bar{F}-a)$&\tiny \begin{tabular}{c}
$-2F(N-(\bar{F}-a))r -2\bar{F}(N-(\bar{F}-a)) \bar{r} $\\$+2F(N-(\bar{F}-a))+2\bar{F}(N-(\bar{F}-a))-2(N^2-(\bar{F}-a)^2)+(\bar{F}-a)(2F-2\bar{F})\bar{r}$
\end{tabular}  \\ \hline
  \end{tabular}}
  \end{center}\label{CBmapping}
\end{table}

Let us consider the matching of the dressed monopole operators. 
From Table \ref{CBmapping} above, the identification becomes
\begin{align}
Y_{a_e}^{dressed} \leftrightarrow \tilde{Y}_{a_m=\bar{F}-a_e}^{dressed}.
\end{align}
The index $a_e$ on the electric side runs over
\begin{align}
 N-(F-\bar{F}) \le a_e  \le \bar{F},~~~0 \le a_e \le N-1.
\end{align}
while the magnetic $a_m$ runs over
\begin{align}
\bar{F}-N \le a_m  \le \bar{F},~~~0 \le a_m \le F-N-1.
\end{align}
Generally, the numbers of the Coulomb branch operators are different for the electric and magnetic sides.
The missing operators appear from the anti-baryonic operators. For $F>N>\bar{F} \cap F-N < \bar{F}$, there are $F-N+1$ electric Coulomb branch operators while there are $F-N$ Coulomb branch operators and one anti-baryon in a magnetic theory. For $F>N>\bar{F} \cap F-N > \bar{F}$, there are $\bar{F}+1$ Coulomb branch operators and the baryons on both sides. For $F>\bar{F}>N \cap F-N < \bar{F}$, there are $F-\bar{F}$ Coulomb branch operators on both sides and (anti-)baryons are available also on both side. Finally, for $F>\bar{F}>N \cap F-N > \bar{F}$, there are $N$ Coulomb branch operators and (anti-)baryons constructible on the electric side, while there are $N+1$ Coulomb branch operators and only a single baryon is possible. In this way, the total numbers of the Coulomb branches and (anti-)baryons are identical and the proposed duality works well.

The precise matching of the Coulomb branch operators (including baryonic ones) is involved and depends on the combination $(N,F,\bar{F})$. Furthermore, depending on $(N,F,\bar{F},a)$, the Coulomb branch operators which have more smaller magnetic charges are available. Then, in what follows, we will consider the specific cases of the duality for the $SU(N)$ ``chiral'' theories.

\subsection{$SU(2N)$ with $(2N+2,2\bar{F})$}
As a first example of our duality, let us start with the case where the dual gauge group becomes $SU(2)$. This case is simplest since the bare monopole operator of the magnetic $SU(2)$ theory is gauge invariant without ``dressing''. The electric theory is a 3d $\mathcal{N}=2$ $SU(2N)$ gauge theory with $(2N+2,2\bar{F})$ (anti-)fundamental matters. The magnetic description is a 3d $\mathcal{N}=2$ $SU(2)$ gauge theory with $2N+2 +2\bar{F}$ fundamentals $(q,\tilde{q})$ and a singlet $M$. The magnetic theory contains the superpotential $W=M q \tilde{q}$. Table \ref{SU2nSU2dual} shows the quantum numbers of the matter contents. 

\begin{table}[H]\caption{3d $\mathcal{N}=2$ $SU(2N)$ with $(2N+2,2\bar{F})$ (anti-)fundamentals} 
\begin{center}
\scalebox{0.88}{
  \begin{tabular}{|c||c||c|c|c|c|c| } \hline
 &$SU(2N)$&$SU(2N+2)$&$SU(2 \bar{F})$&$U(1)$&$U(1)$&$U(1)_R$  \\ \hline
 $Q$ & ${\tiny \yng(1)}$ & ${\tiny \yng(1)}$&1&1&0&$r$ \\
$\tilde{Q}$  &${\tiny \overline{\yng(1)}}$&1& ${\tiny \yng(1)}$&$0$&1&$\bar{r}$ \\  \hline
$M:=\tilde{Q}Q$&1&${\tiny \yng(1)}$& ${\tiny \yng(1)}$&1&1&$r+\bar{r}$ \\ 
$B:=Q^{2N}$&1&${\tiny \overline{\yng(1,1)}}$&1&$2N$&$0$&$2Nr$ \\ \hline
$Y_{\bar{F}-1}^{\substack{minimal \\ dressed}}$&1&1&${\tiny \overline{\yng(1,1)}}$&$-2(N+1)$&$-2$&$4-2(N+1)r -2\bar{r}$ \\
$Y_{\bar{F}}^{dressed} $&1&1&1 & {\scriptsize $-2(N+1)(N-\bar{F})$}&$2\bar{F}$& {\scriptsize $ -2(N+1)(N-\bar{F})r+2\bar{F} \bar{r}+ 2(N-\bar{F}) $ } \\ \hline \hline 
&$SU(2)$&$SU(2N+2)$&$SU(2\bar{F})$&$U(1)$&$U(1)$&$U(1)_R$  \\ \hline
$q$& ${\tiny \yng(1)}$& ${\tiny \overline{\yng(1)}}$&1&$N$&$0$& $ N r$ \\
$\tilde{q}$&${\tiny \yng(1)}$&1 & ${\tiny \overline{ \yng(1)}}$&$-(N+1)$&$-1$&$2-(N+1)r-\bar{r}$ \\
$M$&1&${\tiny \yng(1)}$ & ${\tiny \yng(1)}$&1&1&$r+\bar{r}$ \\\hline
$N:=\tilde{q}q$&1&${\tiny \overline{\yng(1)}}$ & ${\tiny \overline{\yng(1)}}$&$-1$&$-1$&$2-r-\bar{r}$ \\ 
$B:=q^{2} $&1&${\tiny \overline{\yng(1,1)}}$&1&$2N$&$0$&$(2N+1)r$ \\ 
$\bar{B} := \tilde{q}^2 $ &1&1&${\tiny \overline{\yng(1,1)}}$&$-2(N+1)$&$-2$&$4-2(N+1)r -2\bar{r}$ \\ \hline
$\tilde{Y}_{SU(2)}$&1&1&1& {\scriptsize $-2(N+1)(N-\bar{F})$}&$2\bar{F}$& {\scriptsize  $-2(N+1)(N-\bar{F})r+2\bar{F} \bar{r}+ 2(N-\bar{F}) $ }\\ \hline
  \end{tabular}}
  \end{center}\label{SU2nSU2dual}
\end{table}

The bare monopole operators $Y_a^{bare}$ are allowed for $a =\bar{F}, \bar{F}-1$. These two operators are not gauge invariant and require ``dressing'' by $\tilde{Q}^{2a}$. For $a= \bar{F}-1$, the gauge invariant combination can be constructed from $\tilde{Q}^{2a}$ and the minimal bare monopole 
\begin{align}
Y_{a=\bar{F}-1}^{\substack{ minimal \\ bare }}&:=( Y_{\bar{F}-1}^{ bare } )^{\frac{1}{N-\bar{F}+1}},
\end{align}
instead of the naive one $Y_{\bar{F}-1}^{ bare }$. Then, the dressed Coulomb branch will be described by this smaller monopole. For $a= \bar{F}$, the minimal monopole cannot be made gauge invariant by $\tilde{Q}^{2a}$ and we must use $Y_{a=\bar{F}-1}^{bare}$ for constructing the gauge invariant operators. The dressed coordinates become
\begin{align}
Y_{\bar{F}-1}^{\substack{minimal \\ dressed}}&:= Y_{\bar{F}-1}^{\substack{minimal \\ bare}} \tilde{Q}^{2(\bar{F}-1)} \\
Y_{\bar{F}}^{dressed} &:=Y_{\bar{F}}^{bare}  (\tilde{Q}^{2\bar{F}} )^{N+1-\bar{F}}.
\end{align}
In Table \ref{bareCBcharges}, the $U(1)_2$ charges of the Coulomb branch operators are summarized. We can see that $Y_{\bar{F}}^{\substack{minimal \\ bare}}:= (Y_{\bar{F}}^{bare})^{\frac{1}{N-\bar{F}}}$ cannot be dressed by $\tilde{Q}^{2a}$.

\begin{table}[H]\caption{Monopole and dressing operators on the electric side} 
\begin{center}
\scalebox{1}{
  \begin{tabular}{|c||c| }  \hline
  & $U(1)_2$ \\ \hline
  $Y_{\bar{F}-1}^{\substack{minimal \\ bare}}$ & $-(2N+2-2\bar{F})$\\
  $Y_{\bar{F}-1}^{bare}$& $-(2N+2-2\bar{F}) (N-\bar{F}+1)$ \\
  $Q^{2(\bar{F}-1)} |_{a=\bar{F}-1}$ & $2(N-\bar{F}+1)$ \\ \hline
  $Y_{\bar{F}}^{\substack{minimal \\ bare}}$ & $-(2N+2-2\bar{F}) $\\
  $Y_{\bar{F}}^{bare}$& $-(2N+2-2\bar{F}) (N-\bar{F})$ \\
  $Q^{2\bar{F}}|_{a=\bar{F}} $ & $2(N-\bar{F})$\\ \hline
    \end{tabular}}
  \end{center}\label{bareCBcharges}
\end{table}

The operator mapping is manifest from Table \ref{SU2nSU2dual}. Notice that the $SU(2)$ magnetic theory allows only one Coulomb branch operator \cite{Aharony:1997bx} and an additional Coulomb branch comes from the anti-baryonic operator $\bar{B} := \tilde{q}^2 $. The identification becomes 
\begin{gather}
Y_{\bar{F}-1}^{dressed} \sim \bar{B},~~~ Y_{\bar{F}}^{dressed} \sim \tilde{Y}_{SU(2)}.
\end{gather}
For $\bar{F}=N$, the anti-baryon operator $\bar{b}:= \tilde{Q}^{2N}$ is available while the Coulomb branch operator $Y_{\bar{F}=N}$ is not defined. The anti-baryon $\bar{b}$ is identified with the $SU(2)$ monopole operator $\tilde{Y}_{SU(2)}$ in this case. In this way, the duality works even for $\bar{F}=N$.

\subsubsection*{$USp(2(N+\bar{F} -1))$ dual}
Since the dual gauge group $SU(2)$ can be regarded as a member of the $USp(2n)$ series, one can apply the Aharony duality \cite{Aharony:1997gp} to the magnetic side. The Aharony dual is given by a 3d $\mathcal{N}=2$ $USp(2(N+\bar{F} -1))$ gauge theory. The matter contents are summarized in Table \ref{USpdual1}. Since the $USp(2(N+\bar{F} -1))$ theory has only the fundamental matters, the corresponding Coulomb branch is simple and one-dimensional, which is denoted by $Y_{USp}$. 

\begin{table}[H]\caption{$USp(2(N+\bar{F} -1))$ dual of $SU(2N)$ with $(2N+2,2\bar{F})$} 
\begin{center}
\scalebox{0.87}{
  \begin{tabular}{|c||c||c|c|c|c|c| } \hline
 &\scriptsize $USp(2(N+\bar{F}-1))$&\scriptsize $SU(2N+2)$&\scriptsize $SU(2 \bar{F})$&$U(1)$&$U(1)$&$U(1)_R$  \\ \hline
 $b$ & ${\tiny \yng(1)}$ & ${\tiny \yng(1)}$&1&$-N$&0&$1-Nr$ \\
$\tilde{b}$  &${\tiny \yng(1)}$&1& ${\tiny \yng(1)}$&$N+1$&1&$-1+(N+1)r+\bar{r}$ \\  
 $M $&1&${\tiny \yng(1)}$& ${\tiny \yng(1)}$&1&1&$r+\bar{r}$ \\ 
$N\sim q \tilde{q}$&1&${\tiny \overline{\yng(1)}}$&${\tiny \overline{\yng(1)}}$&$-1$&$-1$&$2-r -\bar{r}$ \\
$B \sim q^{2}$&1&${\tiny \overline{\yng(1,1)}}$&1&$2N$&$0$&$2Nr$ \\ 
$\bar{B} \sim \tilde{q}^2$&1&1&${\tiny \overline{\yng(1,1)}}$&$-2(N+1)$&$-2$&$4-2(N+1)r-2\bar{r}$ \\  
$\tilde{Y}_{SU(2)}$&1&1&1&\scriptsize $-2(N+1)(N-\bar{F})$&$2\bar{F}$&\scriptsize $2(N-\bar{F}) -2(N+1)(N-\bar{F})r+2\bar{F} \bar{r}$ \\ \hline
$b \tilde{b} \sim M$&1&${\tiny \yng(1)}$&${\tiny \yng(1)}$&1&1&$r+\bar{r}$ \\
$Y_{USp}$&1&1&1&\scriptsize $2(N+1)(N-\bar{F})$&$-2\bar{F}$&\scriptsize $2(\bar{F}-N+1) +2(N+1)(N-\bar{F})r-2\bar{F} \bar{r}$  \\ \hline 
  \end{tabular}}
  \end{center}\label{USpdual1}
\end{table}

The superpotential takes 
\begin{align}
W &=MN+N b \tilde{b}  +B b^2 +\bar{B}\tilde{b}^2 + \tilde{Y}_{SU(2)}Y_{USp}\\
&=B b^2 +\bar{B}\tilde{b}^2 + \tilde{Y}_{SU(2)}Y_{USp}
\end{align}
Since the mesonic fields $M$ and $N$ are massive, they can be integrated out. The F-term conditions lead to $N=0$ and $M\sim b \tilde{b}$. The matching of the flat directions with the $SU(2N)$ theory is manifest from Table \ref{SU2nSU2dual} and Table \ref{USpdual1}. The unnecessary flat directions are all lifted by the above superpotential.

\subsection{$SU(2N+1)$ with $(2N+3,2\bar{F}+1)$}
For completeness, we study the $SU(2N+1)$ duality whose dual is again given by an $SU(2)$ gauge theory.
The electric theory is a 3d $\mathcal{N}=2$ $SU(2N+1)$ gauge theory with $2N+3$ fundamentals and $2\bar{F}+1$ anti-fundamentals without superpotential. We assume $\bar{F} \le N$ in what follows.

The Coulomb branch operators $Y_a^{bare}$ are available only for $a=\bar{F}$ and $\bar{F}-1$. These are not gauge invariant and need ``dressing''. For $a=\bar{F}-1$, the more minimal operator $(Y_{a=\bar{F}-1}^{bare})^{1/(N-\bar{F}+1)}$ can be dressed by $\tilde{Q}^{2a+1}$ while the $U(1)_2$ charge of $(Y_{a=\bar{F}}^{bare})^{1/(N-\bar{F})}$ cannot be dressed by $\tilde{Q}^{2a+1}$. Therefore, we need to introduce the following Coulomb branch operators
\begin{align}
Y_{\bar{F}-1}^{\substack{minimal \\ dressed}} &:= (Y_{\bar{F}-1}^{bare})^{\frac{1}{N-\bar{F}+1}} \tilde{Q}^{2\bar{F}-1}, \\
Y_{\bar{F}}^{dressed} &:= Y_{\bar{F}}^{bare} (\tilde{Q}^{2\bar{F}+1})^{N-\bar{F}+1},
\end{align}
where the flavor index of $\tilde{Q}^{2a+1}$ is anti-symmetric as it should be. The matter contents and their charges are summarized in Table \ref{SUoddSU2dual} below.

\begin{table}[H]\caption{3d $\mathcal{N}=2$ $SU(2N+1)$ with $(2N+3,2\bar{F}+1)$ (anti-)fundamentals} 
\begin{center}
\scalebox{0.83}{
  \begin{tabular}{|c||c||c|c|c|c|c| } \hline
 &\scriptsize$SU(2N+1)$&\scriptsize$SU(2N+3)$&\scriptsize$SU(2 \bar{F}+1)$&$U(1)$&$U(1)$&$U(1)_R$  \\ \hline
 $Q$ & ${\tiny \yng(1)}$ & ${\tiny \yng(1)}$&1&1&0&$r$ \\
$\tilde{Q}$  &${\tiny \overline{\yng(1)}}$&1& ${\tiny \yng(1)}$&$0$&1&$\bar{r}$ \\  \hline
$M:=\tilde{Q}Q$&1&${\tiny \yng(1)}$& ${\tiny \yng(1)}$&1&1&$r+\bar{r}$ \\ 
$B:=Q^{2N}$&1&${\tiny \overline{\yng(1,1)}}$&1&$2N+1$&$0$&$(2N+1)r$ \\ \hline
$Y_{\bar{F}-1}^{\substack{minimal\\dressed}}$&1&1&${\tiny \overline{\yng(1,1)}}$&\scriptsize$-(2N+3)$& $-2$&$4-(2N+3)r -2\bar{r}$ \\
$Y_{\bar{F}}^{dressed} $&1&1&1 & {\scriptsize $-(2N+3)(N-\bar{F})$}& \scriptsize$2\bar{F}+1$& {\scriptsize $ -(2N+3)(N-\bar{F})r+(2\bar{F} +1)\bar{r}+ 2(N-\bar{F}) $ } \\ \hline \hline 
&\scriptsize$SU(2)$&\scriptsize$SU(2N+3)$&\scriptsize$SU(2\bar{F}+1)$&$U(1)$&$U(1)$&$U(1)_R$  \\ \hline
$q$& ${\tiny \yng(1)}$& ${\tiny \overline{\yng(1)}}$&1&$N+\frac{1}{2}$&$0$& $ \left(N +\frac{1}{2} \right) r$ \\
$\tilde{q}$&${\tiny \yng(1)}$&1 & ${\tiny \overline{ \yng(1)}}$&$- \left(N+ \frac{3}{2} \right)$&$-1$&$2-\left(N+ \frac{3}{2} \right)r-\bar{r}$ \\
$M$&1&${\tiny \yng(1)}$ & ${\tiny \yng(1)}$&1&1&$r+\bar{r}$ \\\hline
$N:=\tilde{q}q$&1&${\tiny \overline{\yng(1)}}$ & ${\tiny \overline{\yng(1)}}$&$-1$&$-1$&$2-r-\bar{r}$ \\ 
$B:=q^{2} $&1&${\tiny \overline{\yng(1,1)}}$&1&$2N+1$&$0$&$(2N+1)r$ \\ 
$\bar{B} := \tilde{q}^2 $ &1&1&${\tiny \overline{\yng(1,1)}}$&$-(2N+3)$&$-2$&$4-(2N+3)r -2\bar{r}$ \\ \hline
$\tilde{Y}_{SU(2)}$&1&1&1& {\scriptsize $-(2N+3)(N-\bar{F})$}&\scriptsize $2\bar{F}+1$& {\scriptsize  $-(2N+3)(N-\bar{F})r+(2\bar{F}+1) \bar{r}+ 2(N-\bar{F}) $ }\\ \hline
  \end{tabular}}
  \end{center}\label{SUoddSU2dual}
\end{table}

On the dual side, the gauge group is $SU(2)$ and there are $2N+2\bar{F}+4$ fundamental matters (no difference between $\mathbf{2}$ and $\bar{\mathbf{2}}$). The dual theory has a superpotential
\begin{align}
W= M \tilde{q}q.
\end{align}
We need not introduce any dressed operators and there is a  single Coulomb branch operator $\tilde{Y}_{SU(2)}$ \cite{Aharony:1997bx}. The operator matching again shows that one of the Coulomb branch operators $Y_{\bar{F}-1}^{\substack{minimal \\ dressed}} $ is transformed into the anti-baryonic operator $\bar{B}$ constructed from the dual quarks $\tilde{q}$.

\subsubsection*{$USp(2(N+\bar{F}))$ dual}
Since the dual gauge group is an $SU(2)$, we can obtain additional dual description by adopting the Aharony duality \cite{Aharony:1997gp}. The $USp(2(N+\bar{F}))$ dual theory contains $2N+2\bar{F}+4$ fundamental matters and singlets $M,N,B,\bar{B}$ and $Y_{USp}$. Table \ref{USpdual2} shows their quantum numbers.

\begin{table}[H]\caption{$USp(2(N+\bar{F} ))$ dual of $SU(2N+1)$ with $(2N+3,2\bar{F}+1)$} 
\begin{center}
\scalebox{0.825}{
  \begin{tabular}{|c||c||c|c|c|c|c| } \hline
 &\scriptsize $USp(2(N+\bar{F}))$&\scriptsize $SU(2N+3)$&\scriptsize $SU(2 \bar{F}+1)$&$U(1)$&$U(1)$&$U(1)_R$  \\ \hline
 $b$ & ${\tiny \yng(1)}$ & ${\tiny \yng(1)}$&1&$- \left( N+\frac{1}{2}\right)$&0&$1-\left( N+\frac{1}{2}\right)r$ \\
$\tilde{b}$  &${\tiny \yng(1)}$&1& ${\tiny \yng(1)}$&$N+\frac{3}{2}$&1&$-1+ \left(N+\frac{3}{2}\right)r+\bar{r}$ \\  
 $M $&1&${\tiny \yng(1)}$& ${\tiny \yng(1)}$&1&1&$r+\bar{r}$ \\ 
$N\sim q \tilde{q}$&1&${\tiny \overline{\yng(1)}}$&${\tiny \overline{\yng(1)}}$&$-1$&$-1$&$2-r -\bar{r}$ \\
$B \sim q^{2}$&1&${\tiny \overline{\yng(1,1)}}$&1&$2N+1$&$0$&$(2N+1)r$ \\ 
$\bar{B} \sim \tilde{q}^2$&1&1&${\tiny \overline{\yng(1,1)}}$&$-(2N+3)$&$-2$&$4-(2N+3)r-2\bar{r}$ \\  
$\tilde{Y}_{SU(2)}$&1&1&1&\scriptsize $-(2N+3)(N-\bar{F})$&\scriptsize $2\bar{F}+1$&\scriptsize $2(N-\bar{F}) -(2N+3)(N-\bar{F})r+(2\bar{F} +1)\bar{r}$ \\ \hline
$b \tilde{b} \sim M$&1&${\tiny \yng(1)}$&${\tiny \yng(1)}$&1&1&$r+\bar{r}$ \\
$Y_{USp}$&1&1&1&\scriptsize $(2N+3)(N-\bar{F})$&\scriptsize $-(2\bar{F}+1)$&\scriptsize $2(\bar{F}-N+1) +(2N+3)(N-\bar{F})r-(2\bar{F}+1) \bar{r}$  \\ \hline 
  \end{tabular}}
  \end{center}\label{USpdual2}
\end{table}

The superpotential takes 
\begin{align}
W&=MN+N b \tilde{b}  +B b^2 +\bar{B}\tilde{b}^2 + \tilde{Y}_{SU(2)}Y_{USp}\\
&=B b^2 +\bar{B}\tilde{b}^2 + \tilde{Y}_{SU(2)}Y_{USp},
\end{align}
where the mesonic fields $M$ and $N$ are massive and can be integrated out. The F-term conditions lead to $N=0$ and $M\sim b \tilde{b}$. The unnecessary flat directions $b^2$, $\tilde{b}^2$ and $Y_{USp}$ are all lifted by the above superpotential.

\subsection{$SU(2N)$ with $(2F+1,1)$}
Let us consider the 3d $\mathcal{N}=2$ $SU(2N)$ gauge theory with $(2F+1,1)$ (anti-)fundamentals. In this case, the Coulomb branch is highly simplified and only $Y_{a=0}^{bare}$ can survive because we cannot anti-symmetrize the anti-fundamental matters. Table \ref{SUevenodd} shows the matter contents and their quantum numbers. The Coulomb brach $Y_{a=0}^{bare}$ is gauge invariant and does not need ``dressing''.

\begin{table}[H]\caption{3d $\mathcal{N}=2$ $SU(2N)$ with $(2F+1,1 )$ (anti-)fundamentals} 
\begin{center}
\scalebox{0.75}{
  \begin{tabular}{|c||c||c|c|c|c| } \hline
 &$SU(2N)$&$SU(2F+1)$&$U(1)$&$U(1)$&$U(1)_R$  \\ \hline
 $Q$ & ${\tiny \yng(1)}$ & ${\tiny \yng(1)}$&1&0&$r$ \\
$\tilde{Q}$  &${\tiny \overline{\yng(1)}}$&1&$0$&1&$\bar{r}$ \\  \hline
$M:=\tilde{Q}Q$&1&${\tiny \yng(1)}$&1&1&$r+\bar{r}$ \\ 
$B:=Q^{2N}$&1&\scriptsize \begin{tabular}{c}$2N$-th \\ antisymmetric\end{tabular}&$2N$&$0$&$2Nr$ \\ \hline
$Y_0^{bare}$&1&1&$-(2F+1)N$&$-N$&\scriptsize $-(2F+1)N r -N\bar{r} +2N(F+1-N)$ \\ \hline \hline
&$SU(2F-2N+1)$&$SU(2F+1)$&$U(1)$&$U(1)$&$U(1)_R$  \\ \hline
$q$& ${\tiny \yng(1)}$& ${\tiny \overline{\yng(1)}}$&$\frac{2N}{2F-2N+1}$&$0$& $ \frac{2N}{2F-2N+1}r$ \\
$\tilde{q}$&${\tiny \overline{\yng(1)}}$&1 &$-\frac{2F+1}{2F-2N+1}$&$-1$&$2-\frac{2F+1}{2F-2N+1}r-\bar{r}$ \\
$M$&1&${\tiny \yng(1)}$ &1&1&$r+\bar{r}$ \\\hline
$N:=\tilde{q}q$&1&${\tiny \overline{\yng(1)}}$&$-1$&$-1$&$2-r-\bar{r}$ \\ 
$B:=q^{2F-2N+1} $&1&\scriptsize \begin{tabular}{c}$2N$-th \\ antisymmetric\end{tabular}&$2N$&$0$&$2Nr$ \\  \hline 
$\tilde{Y}_0^{bare}$&\scriptsize \begin{tabular}{c} $-2F(F-N)$ \\ under $U(1)_2$\end{tabular}&1&$-\frac{(F-N)(2F+1)(2N-1)}{2F-2N+1}$&$F-N$& \tiny \begin{tabular}{c} $-\frac{(F-N)(2F+1)(2N-1)}{2F-2N+1}r+(F-N)\bar{r} $ \\ $+2(F-N)(N-1)$  \end{tabular}  \\
$\tilde{Y}_0^{dressed}:= \tilde{Y}_0^{bare} \tilde{q}^{F}$&1&1& $-(2F+1)N$&$-N$&\scriptsize $-(2F+1)N r -N\bar{r} +2N(F+1-N)$   \\ \hline
  \end{tabular}}
  \end{center}\label{SUevenodd}
\end{table}

The dual description is a 3d $\mathcal{N}=2$ $SU(2F-2N+1)$ gauge theory with $(2F+1,1)$ (anti-)fundamental (dual) matters and a singlet $M$ with the superpotential
\begin{align}
W= M \tilde{q}q
\end{align}
The charges of the dual fields can be fixed from the superpotential and from the matching of the baryon operator as Table \ref{SUevenodd}. The magnetic Coulomb branch is also allowed only for $\tilde{Y}_{a=0}$. Since the dual gauge group is $SU(2F-2N+1)$, $\tilde{Y}_{a=0}$ need ``dressing'' in contrast to the electric side. The matching of the moduli fields are transparent from Table \ref{SUevenodd}.

\subsection{$SU(2N)$ with $(2F,2)$}
The next simple example is a 3d $\mathcal{N}=2$ $SU(2N)$ gauge theory with $(2F,2)$ (anti-)fundamental matters. The dual description is given by a 3d $\mathcal{N}=2$ $SU(2F-2N)$ with $(2F,2)$ (anti-)fundamentals and a singlet $M$. The dual theory includes the superpotential
\begin{align}
W=M \tilde{q}q.
\end{align}
The matter contents and their quantum numbers are summarized in Table \ref{SUeveneven} below.

\begin{table}[H]\caption{3d $\mathcal{N}=2$ $SU(2N)$ with $(2F,2)$ (anti-)fundamentals} 
\begin{center}
\scalebox{0.7}{
  \begin{tabular}{|c||c||c|c|c|c|c| } \hline
  &$SU(2N)$&$SU(2F)$&$SU(2)$&$U(1)$&$U(1)$&$U(1)_R$  \\ \hline
 $Q$ & ${\tiny \yng(1)}$ & ${\tiny \yng(1)}$&1&1&0&$r$ \\
$\tilde{Q}$  &${\tiny \overline{\yng(1)}}$&1& ${\tiny \yng(1)}$&$0$&1& $\bar{r}$ \\  \hline
$M:=\tilde{Q}Q$&1&${\tiny \yng(1)}$& ${\tiny \yng(1)}$&1&1&$r+\bar{r}$\\ 
$B:=Q^{2N}$&1&\tiny \begin{tabular}{c}$2N$-th \\ antisymmetric\end{tabular}&1&$2N$&$0$&$2Nr$ \\ \hline
\footnotesize $Y_0^{bare}$&1&1&1&$-2FN$&$-2N$&\scriptsize $-2FNr -2N\bar{r}+2N(F+1-N)$ \\
\footnotesize $Y_1^{dressed}:=Y_1^{bare} (\tilde{Q}^2)^{F-1} $&1&1&1&$-2F(N-1)$&$2(F-N)$&\scriptsize $-2F(N-1)r+2(F-N)\bar{r}+2(N-1)(F-N)$ \\ \hline \hline
&$SU(2F-2N)$&$SU(2F)$&$SU(2)$&$U(1)$&$U(1)$&$U(1)_R$  \\ \hline
$q$& ${\tiny \yng(1)}$& ${\tiny \overline{\yng(1)}}$&1&$\frac{N}{F-N}$&$0$& $\frac{N}{F-N}r$ \\
$\tilde{q}$&${\tiny \overline{\yng(1)}}$&1 & ${\tiny \yng(1)}$&$-\frac{F}{F-N}$&$-1$&$2-\frac{F}{F-N}r-\bar{r}$ \\
$M$&1&${\tiny \yng(1)}$ &${\tiny \yng(1)}$ &1&1&$r+\bar{r}$ \\\hline
$N:=\tilde{q}q$&1&${\tiny \overline{\yng(1)}}$ & ${\tiny \yng(1)}$ &$-1$&$-1$&$2-r-\bar{r}$ \\ 
$b:=q^{2F-2N} \sim B$&1&\tiny \begin{tabular}{c}$2N$-th \\ antisymmetric\end{tabular}&1&$2N+1$&$0$&$(2N+1)r$ \\ \hline
\footnotesize $\tilde{Y}_0^{bare}$&1&1&1&$-2F(N-1)$&$2(F-N)$&\scriptsize $-2F(N-1)r +2(F-N)\bar{r} +2(N-1)(F-N)$ \\
\footnotesize $\tilde{Y}_1^{dressed}:=\tilde{Y}_1^{bare} (\tilde{q}^2 )^{F-1}$&1&1&1&$-2FN$&$-2N$&\scriptsize $-2FNr-2N\bar{r} +2N(F+1-N)$ \\ \hline
  \end{tabular}}
  \end{center}\label{SUeveneven}
\end{table}

On the electric side, the Coulomb branch is allowed for $Y_{a}^{bare} \,(a=0,1)$. Since the $U(1)_2$ is absent for $a=0$, $Y_{a=0}^{bare}$ is gauge invariant. For $a=1$, $Y_{a=1}^{bare}$ is not gauge invariant and needs ``dressing''. The indices of the dressing factor $\tilde{Q}^2$ is anti-symmetrized and the dressed operator $Y_1^{dressed}:=Y_1^{bare} (\tilde{Q}^2)^{F-1}$ is a flavor singlet. On the dual side, there are also two Coulomb branches $\tilde{Y}_a^{bare}\,(a=0,1)$. $\tilde{Y}_0^{bare}$ is gauge invariant and corresponds to $Y_1^{dressed}$. $\tilde{Y}_1^{bare}$ is not gauge invariant and must be dressed by $\tilde{q}^2$. The dressed operator $\tilde{Y}_1^{dressed}$ can be identified with $Y_0^{bare}$.

For $F=N+1$, the dual gauge group is $SU(2)$ and only  one monopole operator $\tilde{Y}^{bare}_0\sim \tilde{Y}_{SU(2)}$ is available. The remaining Coulomb branch comes from the anti-baryonic operator defined as $\bar{b}:=\tilde{q}^2$. The dual Coulomb branch operator $\tilde{Y}_{SU(2)} $ is identified with $Y_1^{dressed}$. The identification between $Y_0^{bare}$ and $\bar{b}:=\tilde{q}^2$ needs a special care. On the electric side, we have to take the $N$-th root of $Y_0^{bare}$. Then, we find the matching $Y_0^{\substack{ minimal \\ bare }}:= (Y_0^{bare})^{\frac{1}{N}} \sim \bar{b}$.

\subsection{$SU(2N+1)$ with $(2F+1,1)$}
Let us move on to the duality of the 3d $\mathcal{N}=2$ $SU(2N+1)$ gauge theory with $2F+1$ fundamental matters and a single anti-fundamental matter. The dual description is a 3d $\mathcal{N}=2$ $SU(2F-2N)$ gauge theory with $2F+1$ fundamental (dual) matters, a single anti-fundamental (dual) matter and a singlet $M$ with a superpotential
\begin{align}
W=M \tilde{q}q.
\end{align}
The matter contents and their representations are summarized in Table \ref{SUoddeven}.

\begin{table}[H]\caption{3d $\mathcal{N}=2$ $SU(2N+1)$ with $(2F+1,1 )$ (anti-)fundamentals} 
\begin{center}
\scalebox{0.83}{
  \begin{tabular}{|c||c||c|c|c|c| } \hline
 &$SU(2N+1)$&$SU(2F+1)$&$U(1)$&$U(1)$&$U(1)_R$  \\ \hline
 $Q$ & ${\tiny \yng(1)}$ & ${\tiny \yng(1)}$&1&0&$r$ \\
$\tilde{Q}$  &${\tiny \overline{\yng(1)}}$&1&$0$&1&$\bar{r}$ \\  \hline
$M:=\tilde{Q}Q$&1&${\tiny \yng(1)}$&1&1&$r+\bar{r}$ \\ 
$B:=Q^{2N}$&1&\scriptsize \begin{tabular}{c}$(2N+1)$-th \\ antisymmetric\end{tabular}&$2N+1$&$0$&$(2N+1)r$ \\ \hline
\footnotesize $Y_0^{dressed}:=Y_0^{bare}\tilde{Q}^{F}$&1&1& $-(2F+1)N$&$F-N$&\scriptsize $-(2F+1)Nr +(F-N)\bar{r} +2N(F-N)$ \\ \hline  \hline
&$SU(2F-2N)$&$SU(2F+1)$&$U(1)$&$U(1)$&$U(1)_R$  \\ \hline
$q$& ${\tiny \yng(1)}$& ${\tiny \overline{\yng(1)}}$&$\frac{2N+1}{2F-2N}$&$0$& $ \frac{2N+1}{2F-2N}r$ \\
$\tilde{q}$&${\tiny \overline{\yng(1)}}$&1 &$-\frac{2F+1}{2F-2N}$&$-1$&$2-\frac{2F+1}{2F-2N}r-\bar{r}$ \\
$M$&1&${\tiny \yng(1)}$ &1&1&$r+\bar{r}$ \\\hline
$N:=\tilde{q}q$&1&${\tiny \overline{\yng(1)}}$&$-1$&$-1$&$2-r-\bar{r}$ \\ 
$B:=q^{2F-2N} $&1&\scriptsize \begin{tabular}{c}$(2N+1)$-th \\ antisymmetric\end{tabular}&$2N+1$&$0$&$(2N+1)r$ \\  \hline 
$\tilde{Y}_0^{bare}$&1&1&$-(2F+1)N$&$F-N$& \scriptsize$-(2F+1)Nr -(F-N)\bar{r}+2N(F-N) $  \\ \hline
  \end{tabular}}
  \end{center}\label{SUoddeven}
\end{table}

The Coulomb branch operators are allowed only for $a=0$ on both sides. The electric Coulomb branch operator $Y_0^{bare}$ is not gauge invariant and must be dressed by $\tilde{Q}$. On the other hand, the magnetic Coulomb branch $\tilde{Y}_0^{bare}$ is gauge invariant and can be used for a moduli coordinate. From Table \ref{SUoddeven}, we can see the exact matching these two operator under the duality.

\subsection{$SU(2N+1)$ with $(2F+2,2)$}
The final example is a 3d $\mathcal{N}=2$ $SU(2N+1)$ gauge theory with $2F+2$ fundamental and $2$ anti-fundamental matters. The proposed dual description is given by a 3d $\mathcal{N}=2$ $SU(2F-2N+1)$ gauge theory with $2F+2$ fundamental and $2$ anti-fundamental (dual) matters in addition to a singlet $M$. The dual theory has a cubic interaction
\begin{align}
W=M \tilde{q}q.
\end{align}
The matter contents and their quantum numbers are summarized in Table \ref{SUodd2}. Since we consider the region with $F>N$, anti-baryon operators are not available on both sides.

\begin{table}[H]\caption{3d $\mathcal{N}=2$ $SU(2N+1)$ with $(2F+2,2)$ (anti-)fundamentals} 
\begin{center}
\scalebox{0.7}{
  \begin{tabular}{|c||c||c|c|c|c|c| } \hline
  &$SU(2N+1)$&$SU(2F+2)$&$SU(2)$&$U(1)$&$U(1)$&$U(1)_R$  \\ \hline
 $Q$ & ${\tiny \yng(1)}$ & ${\tiny \yng(1)}$&1&1&0&$r$ \\
$\tilde{Q}$  &${\tiny \overline{\yng(1)}}$&1& ${\tiny \yng(1)}$&$0$&1& $\bar{r}$ \\  \hline
$M:=\tilde{Q}Q$&1&${\tiny \yng(1)}$& ${\tiny \yng(1)}$&1&1&$r+\bar{r}$\\ 
$B:=Q^{2N}$&1&\tiny \begin{tabular}{c}$(2N+1)$-th \\ antisymmetric\end{tabular}&1&$2N+1$&$0$&$(2N+1)r$ \\ \hline
\footnotesize $Y_0^{dressed}:=Y_0^{bare}\tilde{Q}^{F}$&1&1& \tiny \begin{tabular}{c}$F$-th \\ symmetric\end{tabular}&$-2(F+1)N$&$F-2N$&\scriptsize $-2(F+1)Nr+(F-2N)\bar{r} +2N(F-N+1)$ \\  \hline \hline
&$SU(2F-2N+1)$&$SU(2F+2)$&$SU(2)$&$U(1)$&$U(1)$&$U(1)_R$  \\ \hline
$q$& ${\tiny \yng(1)}$& ${\tiny \overline{\yng(1)}}$&1&$\frac{2N+1}{2F-2N+1}$&$0$& $\frac{2N+1}{2F-2N+1}r$ \\
$\tilde{q}$&${\tiny \overline{\yng(1)}}$&1 & ${\tiny \yng(1)}$&$-\frac{2F+2}{2F-2N+1}$&$-1$&$2-\frac{2F+2}{2F-2N+1}r-\bar{r}$ \\
$M$&1&${\tiny \yng(1)}$ &${\tiny \yng(1)}$ &1&1&$r+\bar{r}$ \\\hline
$N:=\tilde{q}q$&1&${\tiny \overline{\yng(1)}}$ & ${\tiny \yng(1)}$ &$-1$&$-1$&$2-r-\bar{r}$ \\ 
$b:=q^{2F-2N+1} \sim B$&1&\tiny \begin{tabular}{c}$(2N+1)$-th \\ antisymmetric\end{tabular}&1&$2N+1$&$0$&$(2N+1)r$ \\ \hline
\footnotesize $\tilde{Y}_0^{dressed}:=\tilde{Y}_0^{bare} (\tilde{q} )^{F}$&1&1&\tiny \begin{tabular}{c}$F$-th \\ symmetric\end{tabular}&$-2(F+1)N$&$F-2N$&\scriptsize $-2(F+1)Nr+(F-2N)\bar{r} +2N(F-N+1)$ \\ \hline
  \end{tabular}}
  \end{center}\label{SUodd2}
\end{table}

The electric Coulomb branch is allowed only for $Y^{bare}_{a=0}$, which corresponds to the breaking 
\begin{align}
SU(2N+1) \rightarrow SU(N) \times SU(N) \times U(1)_1 \times U(1)_2.
\end{align}
The $U(1)_2$ charge of the bare operator is $-2FN$. This charge should be canceled by $1_{0,2N} \in \tilde{Q}$. Then the dressed operator becomes $Y_0^{dressed}:=Y^{bare}_{a=0} \tilde{Q}^{F}$. On the magnetic side, the Coulomb branch is allowed only for $a=0$ and the dressed operator becomes $\tilde{Y}_0^{dressed}:=\tilde{Y}_0^{bare} (\tilde{q} )^{F}$. From the charges of these two operators, it is found that these are identified with each other. Notice that, in this dressing procedure, the Coulomb branch operators become $F$-th symmetric representations of the $SU(2)$ flavor symmetry.

\section{Confinement and SUSY breaking}

In this section, we will consider the confinement phases of the 3d $\mathcal{N}=2$ ``chiral'' $SU(N)$ gauge theory with $(F,\bar{F})$ (anti-)fundamentals. (We again assume $F> \bar{F}$.) The appearance of the confinement phase can be regarded as a special limit of our duality proposed in the previous section. 
We will claim that (s-)confinement phases are more ubiquitous than expected from a naive semi-classical analysis of the moduli space of vacua since the dressed monopole operators would not be flavor singlets and then a confining superpotential is available by properly contracting the flavor indices of the meson, baryon and monopole operators. In earlier works \cite{Intriligator:2013lca}, these phases were overlooked due to lack of understanding of the quantum Coulomb branches. 

\subsection{$SU(N)$ with $(F,\bar{F})=(N+1,N-1)$}
In \cite{Aharony:1997bx}, it was discussed that the 3d $\mathcal{N}=2$ $SU(N)$ gauge theory with $(F,\bar{F})=(N,N)$ vector-like matters exhibits s-confinement. Then one can naively guess that the 3d $\mathcal{N}=2$ $SU(N)$ gauge theory with $(F,\bar{F})=(N+1,N-1)$ also shows a similar confinement since this theory differs very little from the former theory and since the total number of the (anti-)fundamental matters does not change. We here claim that this guess is indeed correct.

Let us consider the Coulomb branch labeled by $Y=Y_1\cdots Y_{N-1}$, where $Y_i$ corresponds to the fundamental monopoles. The expectation value for $Y$ breaks the gauge group to $SU(N-2) \times U(1)_1 \times U(1)_2$ and $Y$ is related to a $U(1)_1$ subgroup. Due to the ``chiral'' nature of the theory, the effective Chern-Simons level is induced for $k_{eff}^{U(1)_1 U(1)_2}$ along the RG flow. Therefore, $Y$ is no longer gauge-invariant and charged under the $U(1)_2$ symmetry. The effective CS term is calculated as
\begin{align}
k_{eff}^{U(1)_1 U(1)_2} =F-\bar{F}
\end{align}
and then the $U(1)_2$ charge of $Y$ is $-2$ in the current case. In order to construct the gauge invariant Coulomb branch, we can appropriately combine $Y$ and $\tilde{Q}$ into $Y_d :=Y\tilde{Q}^{N-2}$. As a result, the Coulomb branch is not a flavor singlet. The quantum numbers of the moduli fields are summarized in Table \ref{Scon1}.

\begin{table}[H]\caption{3d $\mathcal{N}=2$ $SU(N)$ with $N+1$ fundamentals and $N-1$ anti-fundamentals} 
\begin{center}
\scalebox{0.82}{
  \begin{tabular}{|c||c|c|c|c|c|c| } \hline
  &$SU(N)$&$SU(N+1)$&$SU(N-1)$&$U(1)_B$&$U(1)_A$&$U(1)_R$  \\ \hline
 $Q$ & ${\tiny \yng(1)}$ & ${\tiny \yng(1)}$&1&1&1&$r$ \\
$\tilde{Q}$  &${\tiny \overline{\yng(1)}}$&1&${\tiny \yng(1)}$&$-1$&1&$\bar{r}$ \\  \hline
$M:=\tilde{Q}Q$&1&${\tiny \yng(1)}$&${\tiny \yng(1)}$&0&2&$r+\bar{r}$ \\ 
$B:=Q^N$&1&${\tiny \overline{\yng(1)}}$&1&$N$&$N$&$Nr$ \\ \hline
$Y:=Y_1 \cdots Y_{N-1}$&$-2$ under $U(1)_2$&1&1&$-2$&$-2N$&$2-(N+1)r-(N-1)\bar{r}$ \\
$Y_d:=Y \tilde{Q}^{N-2}$&1&1&${\tiny \overline{\yng(1)}}$&$-N$&$-N-2$&$2-(N+1)r -\bar{r}$ \\ \hline
  \end{tabular}}
  \end{center}\label{Scon1}
\end{table}

From Table \ref{Scon1}, we see that the low-energy effective degrees of freedom are described by three singlets $M, B$ and $Y_d$ and that the effective superpotential becomes
\begin{align}
W= Y_d M B. \label{sconSUN}
\end{align}
One can check that the parity anomaly matching is satisfied by this confining description. The another check of this phase is that we can derive the same IR description by using the duality discussed in a previous section. Since the dual gauge group is absent, the dual (anti-)quarks $q$ and $\tilde{q}$ are gauge invariant, which are identified with $B$ and $Y_d$ respectively. The dual superpotential takes the same form as \eqref{sconSUN}. This superpotential can be also derived from the 4d s-confinement phase. Let us consider the 4d $\mathcal{N}=1$ $SU(N)$ gauge theory with $N+1$ flavors \cite{Seiberg:1994bz}, which is s-confining and described by the superpotential
\begin{align}
W= -\det M + B^i M_{ia} \bar{B}^a.
\end{align}
By putting the theory on a circle and introducing the real masses for the $SU(2)$ subgroup of the second $SU(N+1)$ flavor symmetry, the electric theory flows to the ``chiral'' $SU(N)$ theory in Table \ref{Scon1}. The effects of the twisted instanton are turned off in this deformation \cite{Aharony:2013dha}. 
On the confined side, the real masses are introduced to the meson and the anti-baryons. By identifying the anti-baryon as the dressed Coulomb branch and integrating out the massive components, the superpotential is correctly reduced to \eqref{sconSUN}.

In what follows, we will deform this confining phase by introducing the complex masses and flow to the 3d $\mathcal{N}=2$ $SU(N)$ gauge theory with $(F,\bar{F})=(N-a,N-a-2)$, where $a \ge 0$.

\subsubsection*{$SU(N)$ with $(F,\bar{F})=(N,N-2)$}
By introducing a complex mass to one pair of (anti-)fundamental matters, we can flow to a 3d $\mathcal{N}=2$ $SU(N)$ gauge theory with $(F,\bar{F})=(N,N-2)$. The quantum numbers of the moduli coordinates are summarized in Table \ref{QDM}. In this case, the flavor indices of $\tilde{Q}^{N-2}$ inside the dressed monopole $Y_d$ is completely anti-symmetrized and $Y_d$ is now a flavor singlet.  

\begin{table}[H]\caption{3d $\mathcal{N}=2$ $SU(N)$ with $N$ fundamentals and $N-2$ anti-fundamentals} 
\begin{center}
\scalebox{0.89}{
  \begin{tabular}{|c||c|c|c|c|c|c| } \hline
  &$SU(N)$&$SU(N)$&$SU(N-2)$&$U(1)_B$&$U(1)_A$&$U(1)_R$  \\ \hline
 $Q$ & ${\tiny \yng(1)}$ & ${\tiny \yng(1)}$&1&1&1&$r$ \\
$\tilde{Q}$  &${\tiny \overline{\yng(1)}}$&1&${\tiny \yng(1)}$&$-1$&1&$\bar{r}$ \\  \hline
$M:=\tilde{Q}Q$&1&${\tiny \yng(1)}$&${\tiny \yng(1)}$&0&2&$r+\bar{r}$ \\ 
$B:=Q^N$&1&1&1&$N$&$N$&$Nr$ \\ \hline
$Y:=Y_1 \cdots Y_{N-1}$&$-2$ under $U(1)_2$&1&1&$-2$&$-2N+2$&$-Nr-(N-2)\bar{r}$ \\
$Y_d:=Y \tilde{Q}^{N-2}$&1&1&1&$-N$&$-N$&$-Nr $ \\ \hline
  \end{tabular}}
  \end{center}\label{QDM}
\end{table}

From Table \ref{QDM}, we can introduce a quantum-deformed constraint between $B$ and $Y_d$
 \begin{align}
B Y_d =1, \label{QC}
\end{align}
while $M$ has no constraint and will become a free field. This phase can be also derived from the previous subsection via a complex mass deformation $\Delta W= m M_{N+1,N-1}$.

\subsubsection*{$SU(N)$ with $(F,\bar{F})=(N-1,N-3)$}
In this case, there is no Coulomb branch operator since we cannot construct the baryonic operator $\tilde{Q}^{N-2}$. By introducing the complex masses for the previous case, one finds no solution for the F-flatness condition of $M$. Therefore, the supersymmetry would be spontaneously broken for a 3d $\mathcal{N}=2$ $SU(N)$ gauge theory with $(F,\bar{F})=(N-a,N-a-2),~a \ge 1$.

\subsection{$SU(2N)$ with $(F,\bar{F})=(2N+1,2\bar{F}+1)$}
We can generalize the confinement phase observed in the previous subsection. The confinement phases appear by taking a special case of the dualities in Section 3. Let us start with the $SU(2N)$ cases. The theory is a 3d $\mathcal{N}=2$ $SU(2N)$ gauge theory with $2N+1$ fundamentals and $2\bar{F}+1$ anti-fundamentals, where we assume $\bar{F} < N$.

In this case, almost all the Coulomb branches are lifted and only $Y_{a=\bar{F}}$, corresponding to the breaking
\begin{align}
SU(2N) \rightarrow SU(N-\bar{F}) \times SU(N-\bar{F}) \times SU(2\bar{F}) \times U(1)_1 \times U(1)_2,
\end{align}
can survive and remain exactly massless. The bare monopole $Y^{bare}_{\bar{F}}$ is not gauge invariant and should be dressed by $\tilde{Q}^{2\bar{F}}$. Now, we have to be careful about the choice of the correct monopole operator since the $U(1)_2$ charge of the minimal monopole 
\begin{align}
Y_{\bar{F}}^{\small \substack{minimal \\bare}} := ( Y^{bare}_{\bar{F}})^{\frac{1}{N-\bar{F}}}
\end{align}
is $-(2N-2\bar{F})$ and this can be dressed by $\tilde{Q}^{2a=2\bar{F}}$. Therefore, the Coulomb branch will be described by the minimal dressed monopole instead of the naive one.
\begin{align}
Y_{\bar{F}}^{\small \substack{minimal \\dressed}} :=( Y^{bare}_{\bar{F}})^{\frac{1}{N-\bar{F}}} \tilde{Q}^{2\bar{F}}
\end{align}
The matter contents and the quantum numbers of the moduli coordinates are summarized in Table \ref{SCONeven}. From a dual theory point of view, $M:=\tilde{Q}Q$ is regarded as a singlet $M$. The baryon operator $B:=Q^{2N}$ is identified with a dual quark $q$ which is now a gauge-singlet. The Coulomb branch operator $Y_{\bar{F}}^{\substack{minimal \\ dressed}}$ corresponds to the dual anti-quark $\tilde{q}$. 
 
\begin{table}[H]\caption{$SU(2N)$ with $(2N+1,2\bar{F}+1)$ (anti-)fundamentals} 
\begin{center}
\scalebox{0.83}{
  \begin{tabular}{|c||c|c|c|c|c|c| } \hline
  &$SU(2N)$&$SU(2N+1)$&$SU(2\bar{F}+1)$&$U(1)$&$U(1)$&$U(1)_R$  \\ \hline
 $Q$ & ${\tiny \yng(1)}$ & ${\tiny \yng(1)}$&1&1&0&$r$ \\
$\tilde{Q}$  &${\tiny \overline{\yng(1)}}$&1&${\tiny \yng(1)}$&$0$&1&$\bar{r}$ \\  \hline
$M:=\tilde{Q}Q$&1&${\tiny \yng(1)}$&${\tiny \yng(1)}$&1&1&$r+\bar{r}$ \\ 
$B:=Q^{2N}$&1&${\tiny \overline{\yng(1)}}$&1&$2N$&$0$&$2Nr$ \\ \hline
\footnotesize $Y_{\bar{F}}^{\substack{minimal \\ dressed}} :=(Y_{\bar{F}}^{bare} )^{\frac{1}{N-\bar{F}}} \tilde{Q}^{2\bar{F}} $&1&1&${\tiny \overline{\yng(1)}}$&$-(2N+1)$&$-1$&$2-(2N+1)r -\bar{r}$  \\  \hline
  \end{tabular}}
  \end{center}\label{SCONeven}
\end{table}

The theory is dual to a non-gauge theory with three chiral superfields $M, B$ and $Y_{\bar{F}}^{\substack{minimal \\ dressed}}$ with a cubic superpotential
\begin{align}
W=BM Y_{\bar{F}}^{\substack{minimal \\ dressed}} =M \tilde{q}q.
\end{align}
This s-confining phase generalizes the result of the previous subsection. By introducing a complex mass, it  is shown that a 3d $\mathcal{N}=2$ $SU(2N)$ gauge theory with $(2N,2\bar{F})$ has a single quantum constraint between the baryon and monopole operators $BY_{\bar{F}}^{\substack{minimal \\ dressed}} =1$ and there is a free meson $M$, which is the same as \eqref{QC}. By further intoducing the complex masses to more flavors, we find that a 3d $\mathcal{N}=2$ $SU(2N)$ gauge theory with $(2N-a,2\bar{F}-a)$ (anti-)fundamentals spontaneously breaks the supersymmetry.

\subsection{$SU(2N+1)$ with $(2N+2,2\bar{F})$}
Finally, we consider the confinement phase for the $SU(2N+1)$ chiral gauge theory. This case also comes from the $SU(2N+1)$ duality by taking the special limit where the dual gauge group becomes null. The electric theory is a 3d $\mathcal{N}=2$ $SU(2N+1)$ gauge theory with $2N+2$ fundamental and $2\bar{F}$ anti-fundamental matters. Since we are interested in a ``chiral'' theory, we require $\bar{F} \le N$ without loss of generality and the following argument is applicable only to this region.

The Coulomb branch $Y_{a=\bar{F}-1} $, corresponding to the breaking 
\begin{align}
SU(2N+1) \rightarrow SU(N-\bar{F}+1) \times SU(N-\bar{F}+1) \times SU(2\bar{F}-1) \times U(1)_1, \times U(1)_2 
\end{align}
can remain flat after including the quantum corrections. Since $Y_{a=\bar{F}-1}^{bare}$ is not gauge invariant, we have to dress the bare monopole operator by $\tilde{Q}^{2a+1}=\tilde{Q}^{2\bar{F}-1}$. 
In this particular matter contents, we can dress the more smaller monopole by using $\tilde{Q}^{2a+1}$. The minimal monopole corresponds to
\begin{align}
Y_{\bar{F}-1}^{ \substack{minimal \\ bare}} :=( Y_{\bar{F}-1}^{bare})^{\frac{1}{N-\bar{F}+1}}
\end{align}
and the dressed minimal operator becomes
\begin{align}
Y_{\bar{F}-1}^{\substack{minimal \\ dressed}} :=Y_{\bar{F}-1}^{ \substack{minimal \\ bare}}\tilde{Q}^{2\bar{F}-1},
\end{align}
which takes an anti-fundamental representation of the $SU(2\bar{F})$ global symmetry. The matter contents and their charges are calculated in Table \ref{SCONodd} below.

\begin{table}[H]\caption{$SU(2N+1)$ with $(2N+2,2\bar{F})$ (anti-)fundamentals} 
\begin{center}
\scalebox{0.81}{
  \begin{tabular}{|c||c||c|c|c|c|c| } \hline
  &$SU(2N+1)$&$SU(2N+2)$&$SU(2\bar{F})$&$U(1)$&$U(1)$&$U(1)_R$  \\ \hline
 $Q$ & ${\tiny \yng(1)}$ & ${\tiny \yng(1)}$&1&1&0&$r$ \\
$\tilde{Q}$  &${\tiny \overline{\yng(1)}}$&1&${\tiny \yng(1)}$&$0$&1&$\bar{r}$ \\  \hline
$M:=\tilde{Q}Q$&1&${\tiny \yng(1)}$&${\tiny \yng(1)}$&1&1&$r+\bar{r}$ \\ 
$B:=Q^{2N+1}$&1&${\tiny \overline{\yng(1)}}$&1&$2N+1$&$0$&$(2N+1)r$ \\ \hline
\footnotesize $Y_{\bar{F}-1}^{\substack{minimal \\ dressed}} :=(Y_{\bar{F}-1}^{bare} )^{\frac{1}{N-\bar{F}+1}} \tilde{Q}^{2\bar{F}-1} $&1&1&${\tiny \overline{\yng(1)}}$&$-(2N+2)$&$-1$&$2-(2N+2)r -\bar{r}$  \\  \hline
  \end{tabular}}
  \end{center}\label{SCONodd}
\end{table}

The low-energy dynamics is described by $M, B$ and $Y_{\bar{F}-1}^{\substack{minimal \\ dressed}}$ with a cubic superpotential
\begin{align}
W=BM Y_{\bar{F}-1}^{\substack{minimal \\ dressed}}.
\end{align}
One can easily check that the parity anomaly matching is satisfied by this dual description. The another check is that one can flow to the $SU(2N)$ s-confinement phase by giving an expectation value to $\braket{M}$.

By introducing the complex masses to this s-confinement phase, we find that a
 3d $\mathcal{N}=2$ $SU(2N+1)$ gauge theory with $(2N+1,2\bar{F}-1)$ (anti-)fundamentals has a quantum-deformed moduli space of the baryon and monopole operators $B Y_{\bar{F}-1}^{\substack{minimal \\ dressed}} =1$ with a single free field $M$. It is also found that the supersymmetry is spontaneously broken for an $SU(2N+1)$ gauge theory with $(2N+1-a,2\bar{F}-1-a)$ (anti-)fundamentals, where $a \ge 1$.
By combining the results of the $SU(2N)$ and $SU(2N+1)$ cases, we conclude that the 3d $\mathcal{N}=2$ $SU(N)$ gauge theory with $(F,\bar{F})$ (anti-)fundamentals exhibits a spontaneous supersymmetry breaking for $N>F>\bar{F}$.

\section{Summary and Discussion}

In this paper, we investigated the quantum structure of the Coulomb moduli space of vacua in the 3d $\mathcal{N}=2$ $SU(N)$ ``chiral'' gauge theories with $F$ fundamental and $\bar{F}$ anti-fundamental matters. In case of a vector-like theory with $F=\bar{F}$, almost all the flat directions of the classical Coulomb branches are lifted and only a single direction ($Y_{N-1}^{bare}$ in our notation) remains exactly massless. On the other hand, in the ``chiral'' $SU(N)$ theories, the very many Coulomb branch directions could generally remain flat, which were described by $Y_a^{bare} (a=0,\cdots,N-1)$. These bare operators are not gauge invariant due to the mixed Chern-Simons term. Then, we constructed the correct gauge-invariant operators by dressing them with the anti-baryonic operators $\tilde{Q}$. Due to this dressing procedure, the Coulomb branch operators are no longer flavor singlets. The previous research \cite{Intriligator:2013lca} was missing this possibility.

By using the ``dressed'' Coulomb branch operators, we proposed the duality for the 3d $\mathcal{N}=2$ $SU(N)$ gauge theory with $(F,\bar{F})$ (anti-)fundamentals. The dual description is given by a 3d $\mathcal{N}=2$ $SU(F-N)$ gauge theory with $(F,\bar{F})$ (anti-)fundamental matters and a meson $M$ with a superpotential $W=M \tilde{q}q$. Naively, one might think that the baryonic operators do not match on both sides of the duality. However, the dressed monopole operators remedy this and the duality works well. The total number of the (anti-)baryonic and Coulomb branch operators is identical on both sides and one can find the correct matching between the Coulomb branch and anti-baryonic operators. 
By focusing on the duality where the dual gauge group is $SU(2)$ and by applying the Aharony duality \cite{Aharony:1997gp}, we found a novel duality between the $SU(N)$ and $USp(2 \tilde{N})$ theories.  
For the 3d $\mathcal{N}=2$ $SU(2)$ gauge theory with $2F$ fundamentals, we can construct the various ``chiral'' dualities. 

We also found the novel confinement phases of the ``chiral'' $SU(N)$ gauge theory. The confinement phases appear for $F=N+1$ and $F>\bar{F}$, where the dual gauge group becomes null. The confined degrees of freedom are the meson, the baryon and the dressed Coulomb branch operator. The low-energy dynamics is governed by the cubic interaction of these three fields whose flavor indices are naturally contracted. By introducing the complex masses to these confining phases, we found that the 3d $\mathcal{N}=2$ $SU(N)$ gauge theory with $(N,N-a)$ (anti-)fundamentals have the free meson $M$ and the quantum deformed moduli space, where $a$ must be even. By further introducing the complex masses, the 3d $\mathcal{N}=2$ $SU(N)$ gauge theory with $(F,\bar{F})$ (anti-)fundamentals exhibits a spontaneous supersymmetry breaking when $N>F>\bar{F}$ and $F+\bar{F}$ is even.

In this paper, we introduced the important class of the Coulomb branch operators $Y_a^{bare}$ which typically appear in the ``chiral'' theories. However, we do not exhaust all the Coulomb branch operators. Generally, there could be other operators. For example, we did not consider the possibility where the dressed monopoles are constructed from the bare monopoles, anti-quarks and gauginos. Typically, they will take the following form
\begin{align}
Y_a^{bare} \tilde{Q}^b \lambda^c, \label{missing}
\end{align}
where $b$ and $c$ are chosen as the above combination becomes gauge invariant.
These operators were studied in case of the $U(N)$ theory \cite{Aharony:2015pla}. One might consider that the matching of these additional operators ruins the proposed duality. However, the matching of the proposed Coulomb branch operators would be highly strong and non-trivial enough to test the validity of the duality. It is worth studying when these additional operators appear or disappear on both sides of the duality. This would be an additional check of our duality.

It would be very important to test our duality and confinement phases from the superconformal index computation \cite{Bhattacharya:2008bja, Kim:2009wb, Imamura:2011su, Kapustin:2011jm}. We have partially done this in Appendix only for the $SU(2)$ case. The additional Coulomb branch operators \eqref{missing} which could appear for more higher rank gauge groups can be studied from the index. Probably, in some cases, the lowest Coulomb branch operator which has a minimal conformal weight could be different from the operators which we defined in this paper. In such cases, \eqref{missing} will give the lowest contribution.
It is also worth understanding our duality and confinement phases from the four-dimensional Seiberg duality point of view by following the same path as \cite{Aharony:2013dha}. This could be achieved by introducing the real masses only for the anti-fundamental matters and flowing to the 3d limit.
The duality of the ``chiral'' $SU(N)$ Chern-Simons matter theories was discussed in \cite{Aharony:2014uya}. The $U(N)$ CS duality was studied in \cite{Giveon:2008zn, Benini:2011mf}. It is curious to find the RG flow from our duality to those Chern-Simons dualities.  


\section*{Acknowledgments}
I would like to thank Ofer Aharony and Antonio Amariti for valuable comments and discussions.
This work is supported by the Swiss National Science Foundation (SNF) under grant number PP00P2\_157571/1 and by ``The Mathematics of Physics'' (SwissMAP) under grant number NCCR 51NF40-141869.

\appendix
\section{$SU(2)$ ``chiral'' and ``non-chiral'' dualities}
In this appendix, we will consider the duality of the 3d $\mathcal{N}=2$ $SU(2)$ gauge theory with $2F$ fundamental matters. This case will further illustrate how our duality works and support the validity of our proposal. 
For concreteness, let us consider the 3d $\mathcal{N}=2$ $SU(2)$ gauge theory with six fundamental matters. The low-energy dynamics of this theory was studied in \cite{Aharony:1997bx, deBoer:1997kr}. In \cite{Dimofte:2012pd} (see also \cite{Benvenuti:2018bav, Amariti:2018wht}), the duality of the $SU(2)$ with six doublets was proposed. Here, we will consider the several ``chiral'' dual descriptions. One of them coincides with the duality in \cite{Dimofte:2012pd}. 

 The Higgs branch is described by a meson composite $M_{QQ}:=QQ$ while the Coulomb branch is parametrized by a single coordinate $Y$. When $Y$ obtains a non-zero vacuum expectation value, the gauge group is broken as $SU(2) \rightarrow U(1)$.
The matter contents and their quantum numbers are summarized in Table \ref{SU(2)F6}. The theory exhibits a manifest global $SU(6)$ symmetry. 

\begin{table}[H]\caption{3d $\mathcal{N}=2$ $SU(2)$ with $6 \, {\tiny \protect\yng(1)} $} 
\begin{center}
\scalebox{1}{
  \begin{tabular}{|c||c||c|c|c| } \hline
  &$SU(2)$&$SU(6)$&$U(1)$&$U(1)_R$  \\ \hline
$Q$&${\tiny \yng(1)}$ &${\tiny \yng(1)}$&1&$r$  \\ \hline
$M_{QQ}:=QQ$&1&${\tiny \yng(1,1)}$&2&$2r$ \\
$Y$&1&1&$-6$&$4-6r$ \\ \hline
  \end{tabular}}
  \end{center}\label{SU(2)F6}
\end{table}

We can regard this theory as the 3d $\mathcal{N}=2$ $SU(2)$ gauge theory with $F$ fundamentals and $6-F$ anti-fundamentals although the explicit $SU(6)$ flavor symmetry is invisible. The Coulomb branch is completely the same as the previous one while the Higgs branch operator is decomposed into the meson, baryon and anti-baryon operators in Table \ref{SU(2)F6expanded}. In the following subsections, we can construct various dual descriptions for each $F$ and some of them are known in the literature \cite{Aharony:1997gp, Dimofte:2012pd, Aharony:2013dha}. 

\begin{table}[H]\caption{3d $\mathcal{N}=2$ $SU(2)$ with $F \, {\tiny \protect\yng(1)} +(6-F) \, {\tiny \overline{\protect\yng(1)}} $} 
\begin{center}
\scalebox{1}{
  \begin{tabular}{|c||c||c|c|c|c|c| } \hline
  &$SU(2)$&$SU(F)$&$SU(6-F)$&$U(1)$&$U(1)$&$U(1)_R$  \\ \hline
$Q$&${\tiny \yng(1)}$ &${\tiny \yng(1)}$&1&1&0&$r$  \\ 
$\tilde{Q}$&${\tiny \overline{\yng(1)}}={\tiny \yng(1)}$&1&${\tiny \yng(1)}$&0&1&$r$ \\ \hline
$M:=QQ$&1&${\tiny \yng(1)}$&${\tiny \yng(1)}$&1&1&$2r$ \\
$B:=Q^2$&1&${\tiny \yng(1,1)}$&1&2&0&$2r$ \\
$\bar{B}:= \tilde{Q}^2$&1&1&${\tiny \yng(1,1)}$&0&2&$2r$ \\
$Y$&1&1&1&$-F$&$F-6$&$4- 6r$ \\ \hline
  \end{tabular}}
  \end{center}\label{SU(2)F6expanded}
\end{table}

\subsection{$USp(2)=SU(2)$ dual}
First, we consider the dual of the description in Table \ref{SU(2)F6} (or $F=6$ in Table \ref{SU(2)F6expanded}). Since the $SU(2)$ group is a member of symplectic groups, we can use the Aharony duality \cite{Aharony:1997gp} and obtain the $USp(2 \tilde{N})$ dual description. The dual gauge group is again $SU(2)$ in this case. The dual theory includes the meson and the Coulomb branch operator as elementary fields. Therefore, all the moduli coordinates are introduced as elementary fields on the dual side. 
\begin{table}[H]\caption{$USp(2)$ dual theory} 
\begin{center}
\scalebox{1}{
  \begin{tabular}{|c||c||c|c|c| } \hline
  &$SU(2)$&$SU(6)$&$U(1)$&$U(1)_R$  \\ \hline
$q$&${\tiny \yng(1)}$ &${\tiny \yng(1)}$&$-1$&$1-r$  \\ 
$M_{QQ}$&1&${\tiny \yng(1,1)}$&2&$2r$ \\
$Y$&1&1&$-6$&$4-6r$ \\ \hline
$\tilde{Y}$&1&1&$6$&$-2+6r$ \\ \hline
  \end{tabular}}
  \end{center}\label{USpSU(2)dual}
\end{table}
\noindent Table \ref{USpSU(2)dual} shows the matter contents and their quantum numbers of the $USp(2)$ dual theory. The dual theory has a tree-level superpotential
\begin{align}
W=M_{QQ}qq + Y \tilde{Y},
\end{align}
which is consistent with all the symmetries in Table \ref{USpSU(2)dual}. The equations of motion of the superpotential lift the magnetic meson and the dual Coulomb branch operators.

\subsection{$SU(2)$ ``chiral'' dual}
Next, we consider the dual of the description in Table \ref{SU(2)F6expanded} for $F=4$. The low-energy dynamics of the ``chiral'' $SU(N)$ theories and the dual description is given in this paper. The dual gauge group again becomes $SU(2)$ and this case coincides with \cite{Dimofte:2012pd}. The dual theory contains four fundamentals, two anti-fundamentals and a meson singlet $M$. The dual theory has a tree-level superpotential
\begin{align}
W=Mq \tilde{q}.
\end{align}
\begin{table}[H]\caption{``Chiral'' $SU(2)$ dual theory} 
\begin{center}
\scalebox{1}{
  \begin{tabular}{|c||c||c|c|c|c|c| } \hline
  &$SU(2)$&$SU(4)$&$SU(2)$&$U(1)$&$U(1)$&$U(1)_R$  \\ \hline
$q$&${\tiny \yng(1)}$ &${\tiny \overline{\yng(1)}}$&1&$1$&0&$r$  \\ 
$\tilde{q}$&${\tiny \overline{\yng(1)}}={\tiny \yng(1)}$&1&${\tiny \yng(1)}$&$-2$&$-1$&$2-3r$ \\ 
$M$&1&${\tiny \yng(1)}$&${\tiny \yng(1)}$&1&1&$2r$ \\ \hline
$B:=q^2$&1&${\tiny \yng(1,1)}$&1&2&0&$2r$ \\
$Y \sim \tilde{q}^2$&1&1&1&$-4$&$-2$&$4-6r$ \\ \hline
$\bar{B} \sim Y_{SU(2)}$&1&1&1&$0$&$2$&$2r$ \\ \hline
  \end{tabular}}
  \end{center}\label{SecondSU(2)dual}
\end{table}

\noindent The quantum numbers of the chiral superfields are summarized in Table \ref{SecondSU(2)dual}. The matching of the moduli operators are as follows. 
\begin{gather}
Q \tilde{Q} \sim  M, ~~~Q^2 \sim q^2 \nonumber \\
\tilde{Q}^2 \sim Y_{SU(2)},~~~Y \sim  \tilde{q}^2
\end{gather}
As we explained in Section 3, the role of the Coulomb branch and the anti-baryonic operator is exchanged. In this dual, only the meson is introduced as an elementary field on the magnetic side. Therefore, the duality has no problem of the UV-completion.  

\subsection{$SU(2)$ third dual}
We can construct the third $SU(2)$ dual description \cite{Dimofte:2012pd} where the (anti-)baryonic operators and the Coulomb branch are introduced as elementary fields. Although this is not a ``chiral'' duality studied in this paper, we will show this duality for completeness.  

\begin{table}[H]\caption{Third $SU(2)$ dual theory} 
\begin{center}
\scalebox{1}{
  \begin{tabular}{|c||c||c|c|c|c|c| } \hline
  &$SU(2)$&$SU(4)$&$SU(2)$&$U(1)$&$U(1)$&$U(1)_R$  \\ \hline
$q$&${\tiny \yng(1)}$ &${\tiny \yng(1)}$&1&$-1$&0&$1-r$  \\ 
$\tilde{q}$&${\tiny \yng(1)}$&1&${\tiny \yng(1)}$&$2$&$1$&$-1+3r$ \\ 
$B$&1&${\tiny \yng(1,1)}$&1&2&0&$2r$ \\
$\bar{B}$&1&1&1&$0$&$2$&$2r$ \\ 
$Y$&1&1&1&$-4$&$-2$&$4-6r$ \\ \hline
$M \sim q \tilde{q}$&1&${\tiny \yng(1)}$&${\tiny \yng(1)}$&1&1&$2r$ \\ \hline 
$\tilde{Y}_{SU(2)}$&1&1&1&$0$&$-2$&$2-2r$ \\ \hline
  \end{tabular}}
  \end{center}\label{ThirdSU(2)dual}
\end{table}
\noindent The matter contents and their quantum numbers are summarized in Table \ref{ThirdSU(2)dual}. The third dual description includes the tree-level superpotential
\begin{align}
W=B q^2 +Y \tilde{q}^2 +\bar{B} \tilde{Y}_{SU(2)},
\end{align}
which is consistent with all the symmetries in Table \ref{ThirdSU(2)dual}. The magnetic meson is mapped to the electric meson operator.

\subsection{$SU(4)$ ``chiral'' dual}
Let us consider the chiral dual description of Table \ref{SU(2)F6expanded} with $F=6$. This case was studied in \cite{Aharony:2013dha}. Since the $SU(2)$ gauge theory with six fundamentals can be regarded as the $SU(2)$ gauge theory with six fundamentals and no anti-fundamental ($F=6$ in Table \ref{SU(2)F6expanded}), we can construct the ``chiral'' $SU(4)$ dual in addition to the $USp(2)$ dual. The dual theory has no tree-level superpotential. The magnetic baryon operator is identified with the electric meson. The Coulomb branch operator $Y$ is mapped to the magnetic Coulomb branch operator $\tilde{Y}_{SU(2) \times SU(2)}$ whose expectation value leads to the breaking $SU(4) \rightarrow SU(2) \times SU(2) \times U(1)$. The quantum numbers of the magnetic matter contents are summarized in Table \ref{SU(4)dual}.

\begin{table}[H]\caption{$SU(4)$ dual of $SU(2)$ with $6\, {\tiny \protect\yng(1)}$} 
\begin{center}
\scalebox{1}{
  \begin{tabular}{|c||c||c|c|c| } \hline
  &$SU(4)$&$SU(6)$&$U(1)$&$U(1)_R$  \\ \hline
$q$&${\tiny \yng(1)}$ &${\tiny  \overline{\yng(1)}}$&$\frac{1}{2}$&$\frac{1}{2}r$  \\ \hline
$q^4$&1&${\tiny \yng(1,1)}$&2&$2r$ \\
$\tilde{Y}_{SU(2) \times SU(2)}$&1&1&$-6$&$4-6r$ \\ \hline
  \end{tabular}}
  \end{center}\label{SU(4)dual}
\end{table}

\subsection{$SU(3)$ ``chiral'' dual}
By regarding the electric theory as the $SU(2)$ gauge theory with five fundamentals and one anti-fundamental ($F=5$ in Table \ref{SU(2)F6expanded}), we can also construct the $SU(3)$ dual description. 
The dual side becomes the 3d $\mathcal{N}=2$ theory with five fundamentals, one anti-fundamental and a gauge singlet $M$. The quantum numbers of these fields are summarized in Table \ref{SU(4)dual}. The dual side includes a tree-level superpotential
\begin{align}
W=Mq \tilde{q},
\end{align}
which is consistent with all the symmetries. The bare Coulomb branch $\tilde{Y}^{bare}$ corresponds to the breaking $SU(3) \rightarrow U(1) \times U(1)$ and needs ``dressing''. The dressed operator $\tilde{Y}^{dressed}:=\tilde{Y}^{bare} \tilde{q}^2$ is mapped to the electric Coulomb branch operator. The magnetic baryon $q^3$ and the meson singlet $M$ are identified with the decomposed electric meson operator. 

\begin{table}[H]\caption{$SU(3)$ dual of $SU(2)$ with $6\, {\tiny \protect\yng(1)}$} 
\begin{center}
\scalebox{1}{
  \begin{tabular}{|c||c||c|c|c|c| } \hline
  &$SU(3)$&$SU(5)$&$U(1)$&$U(1)$&$U(1)_R$  \\ \hline
$q$&${\tiny \yng(1)}$&${\tiny  \overline{\yng(1)}}$&$\frac{2}{3}$&0&$\frac{2}{3}r$ \\
$\tilde{q}$&${\tiny  \overline{\yng(1)}}$&1&$-\frac{5}{3}$&$-1$&$2-\frac{8}{3}r$  \\
$M$&1&${\tiny \yng(1)}$&1&1&$2r$ \\ \hline
$B \sim q^3$&1&${\tiny \yng(1,1)}$&2&0&$2r$ \\  \hline
$\tilde{Y}^{bare}:= \tilde{Y}_1 \tilde{Y}_2$&\footnotesize $U(1)_2$ charge: $-4$&1&$-\frac{5}{3}$&1&$-\frac{2}{3}r$  \\
$\tilde{Y}^{dressed}:=\tilde{Y}^{bare} \tilde{q}^2$&1&1&$-5$&$-1$&$4-6r $ \\  \hline
  \end{tabular}}
  \end{center}\label{SU(4)dual}
\end{table}

\subsection{$U(1)$ ``vector-like'' dual}
Finally, we show the magnetic $U(1)$ dual description which was found in \cite{Aharony:2013dha}. The $SU(2)$ gauge theory with six fundamentals can be regarded as the $SU(2)$ gauge theory with three fundamentals and three anti-fundamentals ($F=3$ in Table \ref{SU(2)F6expanded}). By following \cite{Aharony:2013dha}, we can construct the $U(F-N)$ dual description. The magnetic dual is the 3d $\mathcal{N}=2$ $U(1)$ gauge theory with four electrons $q, \tilde{b}$ and four positrons $\tilde{q}, b$ in addition to the gauge singlets $M$ and $Y$. The quantum numbers of the dual fields are summarized in Table \ref{U(1)dual}.
The theory includes the tree-level superpotential 
\begin{align}
W=M \tilde{q}q +Y b \tilde{b} + \tilde{X}_+  +  \tilde{X}_-,
\end{align}
which is consistent with all the symmetries in Table \ref{U(1)dual}. One can easily find the operator matching from Table \ref{U(1)dual}.

\begin{table}[H]\caption{$U(1)$ ``vector-like'' dual theory} 
\begin{center}
\scalebox{1}{
  \begin{tabular}{|c||c||c|c|c|c|c| } \hline
  &$U(1)$&$SU(3)$&$SU(3)$&$U(1)$&$U(1)$&$U(1)_R$  \\ \hline
$q$&1&${\tiny \overline{\yng(1)}}$&1&$-\frac{1}{2}$&$-\frac{1}{2}$&$1-r$  \\ 
$\tilde{q}$&$-1$&1&${\tiny \overline{\yng(1)}}$&$-\frac{1}{2}$&$-\frac{1}{2}$&$1-r$ \\ 
$b$&$-1$&1&1&$\frac{5}{2}$&$\frac{1}{2}$&$-1 +3r$ \\
$\tilde{b}$&$1$&1&1&$\frac{1}{2}$&$\frac{5}{2}$& $-1 +3r$ \\
$M $&1&${\tiny \yng(1)}$&${\tiny \yng(1)}$&1&1&$2r$ \\
$Y$&1&1&1&$-3$&$-3$&$4-6r$ \\ \hline
$B \sim qb$&1&${\tiny \overline{\yng(1)}}$&1&2&0&$2r$ \\
$\bar{B} \sim \tilde{q} \tilde{b}$&1&1&${\tiny \overline{\yng(1)}}$&$0$&$2$&$2r$ \\  \hline
$\tilde{X}_{ \pm }$&1&1&1&$0$&$0$&$2$ \\ \hline
  \end{tabular}}
  \end{center}\label{U(1)dual}
\end{table}

As a consistency check for these seven descriptions (one electric and six duals), we can compute the superconformal indices. We found that these seven descriptions show a beautiful agreement of the indices. The index is computed as

\scriptsize
\begin{align}
I&=1+\left(\frac{1}{t^6}+15 t^2\right) x+\left(\frac{1}{t^{12}}+105 t^4-36\right) x^2+\left(\frac{1}{t^{18}}+490 t^6-384 t^2+\frac{21}{t^2}\right) x^3+\left(\frac{1}{t^{24}}+1764 t^8-2100 t^4+\frac{36}{t^4}+558\right) x^4 \nonumber \\
&+\left(\frac{1}{t^{30}}+5292 t^{10}-8064 t^6-\frac{71}{t^6}+3690 t^2-\frac{384}{t^2}\right) x^5+\left(\frac{1}{t^{36}}+13860 t^{12}-24696 t^8+\frac{36}{t^8}+14526 t^4+\frac{135}{t^4}-3536\right) x^6+\cdots,
\end{align}
\normalsize
where $t$ is a fugacity for the global $U(1)$ symmetry in Table \ref{SU(2)F6} and we set the r-charge to $r=\frac{1}{2}$ for simplicity. The first term corresponds to the identity operator. The second term consists of the Coulomb branch $Y_{SU(2)}$ and the meson operator $M$. The higher order terms include the fermion contributions and the symmetric products of the moduli operators. 

As we explicitly explained the various chiral dualities for the $SU(2)$ with six doublets, one can easily generalize this argument to the $SU(2)$ gauge theory with $2F$ doublets. By regarding the theory as the $SU(2)$ gauge theory with $2F$ fundamentals and no anti-fundamental, the $USp$ dual and the ``chiral'' $SU(2F-2)$ dual are available. For the $SU(2)$ gauge theory with $2F-a$ fundamentals and $a$ anti-fundamentals, the ``chiral'' $SU(2F-a-2)$ dual theory, which we proposed in this paper, can be constructed. For $a=F$, we can also have the ``vector-like'' dual \cite{Aharony:2013dha}. In these dual descriptions, the full global symmetry is not manifest and only visible in the far-infrared limit.

\bibliographystyle{ieeetr}
\bibliography{3dchiral}

\end{document}